\documentclass[pra,twocolumn,showpacs,superscriptaddress,floatfix]{revtex4}
\usepackage{latexsym,amsfonts,amssymb}
\usepackage{graphicx}

\def\F{{\cal F}}

\def\E{{\cal E}}
\def\RR{\mathbb{R}}

\def\eps{\epsilon}
\def\vec#1{{\bf #1}}
\def\op#1{\hat{#1}}
\def\ket#1{| #1 \rangle}

\def\bra#1{\langle #1 |}

\def\norm#1{\| #1 \|}

\def\Tr{\mathop{\rm Tr}}

\def\diag{\mathop{\rm diag}}
\def\floor#1{\left\lfloor #1 \right\rfloor}
\def\ceil#1{\left\lceil #1 \right\rceil}
\newif\ifpdflatex\pdflatextrue
\makeatletter\@ifundefined{pdfoutput}{\pdflatexfalse}\makeatother
\def\myincludegraphics[#1]#2#3{%
    \ifpdflatex \includegraphics[#1]{#2}
    \else       \includegraphics[#1]{#3}
    \fi}
\ifpdflatex 
    \usepackage[pdftex,backref,colorlinks]{hyperref}
    \hypersetup{%
      pdfauthor  = {Sonia G Schirmer},
      pdftitle   = {Controlled phase gate for solid-state charge-qubit architectures},
      pdfsubject = {quantum computing, quantum control},
      pdfkeywords= {controlled phase gate, charge qubits, entanglement}}
\fi
\begin{document}
\title{Controlled phase gate for solid-state charge-qubit architectures}
\author{S.~G.~Schirmer}
\email{sgs29@cam.ac.uk}
\affiliation{Department of Applied Mathematics and Theoretical Physics,
University of Cambridge, Wilberforce Road, Cambridge CB3 0WA, UK}
\affiliation{Department of Engineering, University of Cambridge,
Trumpington Street, Cambridge, CB2 1PZ, UK}
\author{D.~K.~L.~Oi}
%\email{dklo2@cam.ac.uk}
\affiliation{Department of Applied Mathematics and Theoretical Physics,
University of Cambridge, Wilberforce Road, Cambridge CB3 0WA, UK}
\author{Andrew~D.~Greentree} 
\affiliation{Centre for Quantum Computer Technology, School of Physics, 
University of Melbourne, Melbourne, Victoria 3010, Australia}
\date{\today}
\begin{abstract}
We describe a mechanism for realizing a controlled phase gate for solid-state
charge qubits.  By augmenting the positionally defined qubit with an auxiliary
state, and changing the charge distribution in the three-dot system, we are able
to effectively switch the Coulombic interaction, effecting an entangling gate.
We consider two architectures, and numerically investigate their robustness to 
gate noise.
\end{abstract}
\maketitle

%%%%%%%%%%%%%%%%%%%%%%
\section{Introduction}
%%%%%%%%%%%%%%%%%%%%%%

The search for a workable quantum information processor is an effort that has 
captivated the attention of researchers in many disciplines.  A quantum computer
requires individual quantum logic elements, usually qubits, and entangling 
interactions between these elements~\cite{SCI270p255,PRA52p3457}.

Solid-state proposals are widely seen as being some of the most attractive from 
the point of view of constructional scalability, i.e., the ability to replicate 
many qubits.  Furthermore, schemes compatible with present semiconductor 
technologies~\cite{PRA57p0120,NAT393p133,PRA62n012306,PRB69n113301} are especially
attractive because of their potential to leverage the associated industrial 
experience~\cite{PTRSA361p1451,JAP94p7017}.  

In this paper we concentrate on charge-based architectures.  Such systems were
amongst the first proposed for quantum computing~\cite{RMP68p733} and numerous
versions have evolved recently~\cite{NanoTech11p387, PRB69n113301,PRL91n226804,
NAT398p786}.  We are attracted to charge-based systems for three reasons: 
(1) proven high-fidelity readout compatible with single-shot operations~\cite{NAT406p1039}; 
(2) potential for high-speed ($\sim$ picosecond) operations~\cite{PRB69n113301}; 
and (3) the ability to define variable dimensionality Hilbert spaces by appropriate 
partitioning~\cite{PRL92n097901}.
The usual mechanism for coupling charge qubits is via the Coulomb interaction.
In general, this interaction cannot be controlled without a variation in the 
charge distribution in the qubits. In this paper we specifically address this 
issue and show how to make a scalable controlled phase gate that makes use of
the Coulomb interaction.

%~\cite{PRL79p2371}
%NAT398p786, PRB66n235303, PRL92n098301, 
%PRA66n042328, PRB69n113301, PE21p1046.  

The Coulomb interaction is insensitive to minor variations in the distance 
between the quantum dots and its strength lends itself to high-speed entangling 
operations; conversely, its long-range nature makes it difficult to effectively 
modulate interactions between qubits.  Most charge qubit schemes so far proposed
implicitly rely on a fixed, always-on Coulomb interaction between 
qubits~\cite{PRB69n113301, NanoTech11p387,NAT421p823}  Such schemes would
necessarily require global control techniques~\cite{SCI261p1569,PRL88n017904,
PRL90n247901}, which may be problematic given the strength of the Coulomb 
interaction.

In the following we first discuss local qubit operations (Sec.~\ref{sec:local}) 
and introduce the two-qubit interaction Hamiltonian (Sec.~\ref{sec:2qubitH}).  
Then we describe how to realize a controlled phase gate and analyze its operation
in terms of dynamic and geometric phases (Sec.~\ref{sec:phaseGate}).  Finally, we
discuss various practical issues such as the detrimental effects of finite rise 
and decay times on the gate fidelity and how to correct them (Sec.~\ref{sec:errors}), 
gate implementation in the presence of practical constraints on pulse lengths and 
tunnelling rates (Sec.~\ref{sec:constraints}), and the effect of noisy controls 
(Sec.~\ref{sec:noise}) and imperfect architectures (Sec.~\ref{sec:perturb}) on 
the gate performance.

%%%%%%%%%%%%%%%%%%%%%%%%%%%%%%%%%%
\section{Single Qubit System}
\label{sec:local}
%%%%%%%%%%%%%%%%%%%%%%%%%%%%%%%%%%

\begin{figure}
\myincludegraphics[]{figures/pdf/potential.pdf}{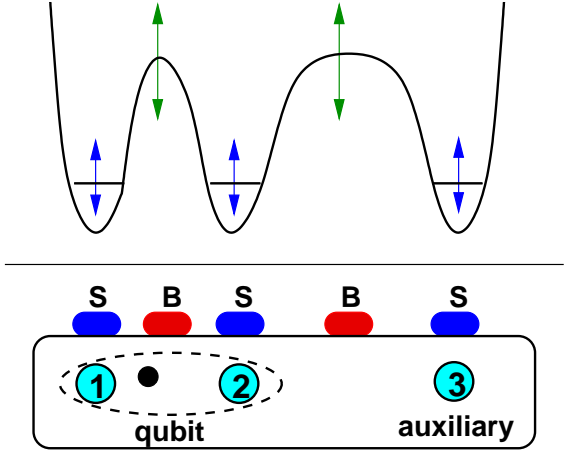}
\caption{Schematic of a single qubit-plus-auxiliary unit (bottom) and electronic
potential (top).  The large circles represent the quantum dots, the black dot the
shared electron.   $S$ and $B$ surface electrodes serve to shift the ground state 
energy of the dots and change the height of the tunnelling barriers and thus the 
tunnelling rates between them.}
\label{fig:potential}
\end{figure}

As will be shown below, a three-dot system is required to modulate the Coulomb
interaction.  We therefore supplement the canonical charge qubit with an 
auxiliary state and define a quantum element with three quantum dots and one 
charge, where the position of the charge on two sites defines the qubit and the 
third site defines an auxiliary state used for two-qubit interactions.  
We work only with the ground state of each dot.  We further assume that the ground
states are energetically close and sufficiently separated from higher-lying
excited states, so that the excitation of these states can be neglected.  The 
system can thus be approximated as a three level system with Hamiltonian
\begin{equation} \label{eq:Hloc1}
  \op{H} = \sum_{d=1}^3\eps_d \ket{d}\bra{d}+\hbar\sum_{d'\neq d}\mu_{dd'}\op{X}_{dd'},
\end{equation}
where $\ket{d}$ denotes the ground state of the electron localized in dot $d$ for 
$d=1,2,3$, $\eps_d$ the energy of state $\ket{d}$, $\mu_{dd'}$ the tunnelling rate
between dots $d$ and $d'$ (for $d'\neq d$), and $\op{X}_{dd'}=\ket{d}\bra{d'}+
\ket{d'}\bra{d}$.  We shall assume that we can control the ground state energies 
$\eps_d$ and the tunnelling rates $\mu_{dd'}$, e.g., by varying the voltages applied
to several control electrodes as illustrated in Fig.~\ref{fig:potential}. 

To implement local unitary operations, we inhibit tunnelling to the auxiliary site
by raising the barrier between dots $2$ and $3$ (and/or $1$ and $3$ if applicable), 
or increasing the ground state energy of the auxiliary dot $3$.  In practice, the 
precise functional dependence of $\eps_d$ and $\mu_{dd'}$ on the control voltages 
applied should be determined experimentally using Hamiltonian identification 
techniques~\cite{PRA69n050306}.  When $\mu_{13}=\mu_{23}=0$ the local Hamiltonian 
can be rewritten as
\begin{equation}
  \op{H} = \eps_1 \ket{1}\bra{1}+\eps_2 \ket{2}\bra{2}+\hbar\mu_{12}\op{X}_{12} 
          +\eps_3\ket{3}\bra{3}, \label{eq:Hloc2}
\end{equation}
and we can realize arbitrary unitary operations on the qubit subspace by changing 
the energy levels $\eps_d$ and the tunnelling rate $\mu_{12}$.  For instance, 
shifting the energy level of dot 2 by $\eps_2(t)$ for $t_0\le t\le t_1$ results 
in a local phase rotation
\begin{equation}
   \op{U}_2(\phi) = \ket{1}\bra{1} + e^{-i\phi}\ket{2}\bra{2} + \ket{3}\bra{3},
\end{equation}
with $\phi=\int_{t_0}^{t_1} \eps_2(t)/\hbar\,dt$.  Effecting a tunnelling rate
of $\mu_{12}$ for $t_0\le t\le t_1$ gives
\begin{equation}
   \op{U}_{12}(\alpha) = \cos(\alpha) \op{I}_{12} -i \sin(\alpha)\op{X}_{12}
                               + \ket{3}\bra{3},
\end{equation}
where the rotation angle is $\alpha=\int_{t_0}^{t_1} \mu_{12}^{(k)}(t)\,dt$ and
$\op{I}_{12}=\ket{1}\bra{1}+\ket{2}\bra{2}$.  By combining two phase rotations on
the 2nd dot with a tunnelling gate between sites~$1$ and $2$, for example, we can
implement any local unitary operation on the qubit subspace modulo global 
phases~\cite{PRA52p3457}
\begin{eqnarray}
  \op{U}(\phi_1,\alpha,\phi_2)
  &=& \op{U}_2(\phi_2)\op{U}_{12}(\alpha)\op{U}_2(\phi_1), \nonumber\\
  &=& \left( \begin{array}{rr|r} 
        \cos(\alpha)                & -ie^{-i\phi_1} \sin(\alpha)       & 0 \\
        -ie^{-i\phi_2} \sin(\alpha) & e^{-i(\phi_1+\phi_2)}\cos(\alpha) & 0 \\ \hline
                                  0 &                                 0 & 1 
       \end{array} \right), \nonumber \\
%   &=& \cos(\alpha) (\ket{1}\bra{1} + e^{-i(\phi_1+\phi_2)}\ket{2}\bra{2})
%       +\ket{3}\bra{3} \nonumber \\
%   & & -i\sin(\alpha) (e^{-i\phi_1} \ket{1}\bra{2}+e^{-i\phi_2} \ket{2}\bra{1}) 
   \label{eq:Ulocal}
\end{eqnarray}
For example, a Hadamard transform on the qubit subspace corresponds to
\begin{equation}
 \op{H} = \frac{1}{\sqrt{2}} \left(\begin{array}{rr} 
                                   1 & 1 \\ 1 & -1 
                             \end{array} \right)
        = \op{U}\left(\phi_2=-\frac{\pi}{2},\alpha=\frac{\pi}{4},
                      \phi_1=-\frac{\pi}{2}\right).
\end{equation}
It is possible to optimize the implementation of some local unitary operations by 
simultaneously changing multiple control parameters.  See appendix \ref{app:local}.

%%%%%%%%%%%%%%%%%%%%%%%%%%%%%%%%%
\section{Interaction Hamiltonian}
\label{sec:2qubitH}
%%%%%%%%%%%%%%%%%%%%%%%%%%%%%%%%%

To achieve entangling operations it is necessary to change the charge distribution
of adjacent elements.  Although almost any interaction will lead to entanglement%
~\cite{PRL75p0346}, the resulting dynamics may be neither easy to utilize, nor 
robust against noise.

\begin{figure*}
\begin{tabular}{cc}
{\sf\bf(a)} \myincludegraphics[]{figures/pdf/arch2D.pdf}{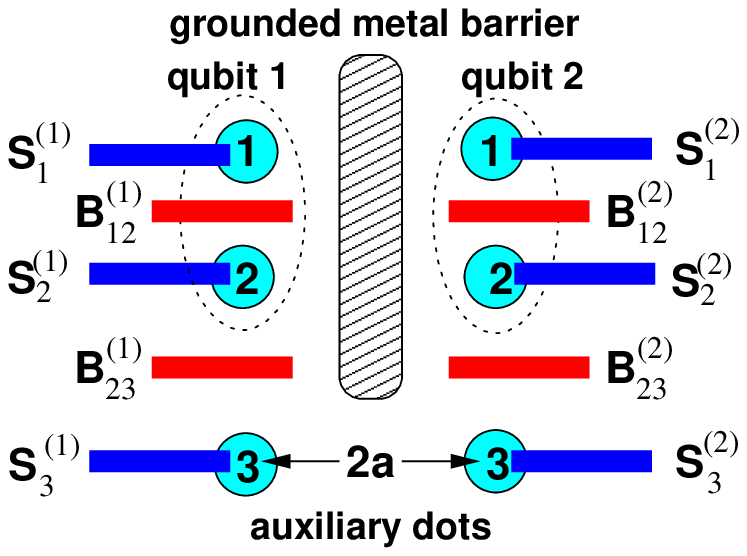} &
{\sf\bf(b)} \myincludegraphics[width=3in]{figures/pdf/arch3D.pdf}{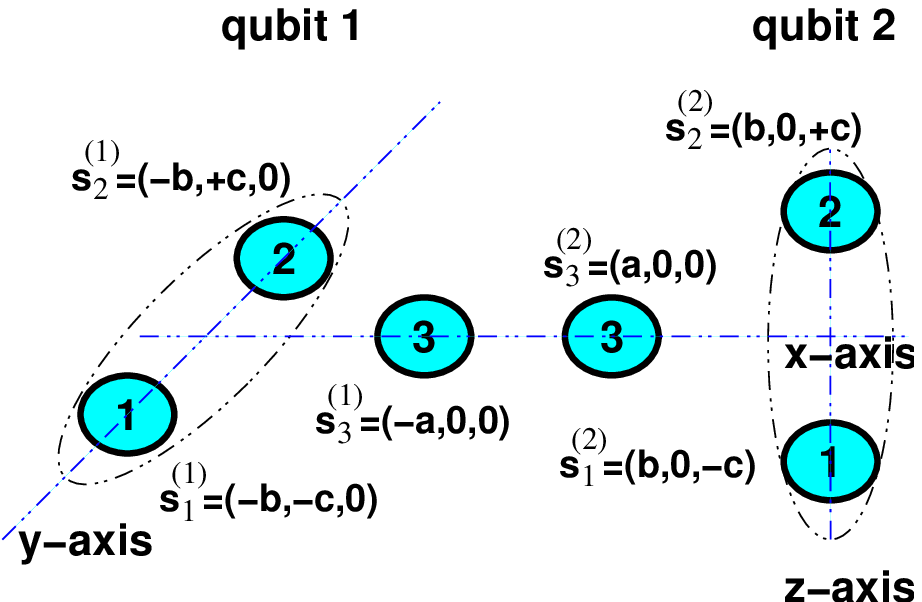}
\end{tabular}
\caption{(a) Two-qubit system with two auxiliary sites comprised of six quantum 
dots (filled circles).  A grounded metal barrier shields the Coulomb interaction
between nearby quantum dots except between the two auxiliary sites (3).  The thick
(red and blue) lines indicate the surface control electrodes $B_{dd'}^{(k)}$ and
$S_{d}^{(k)}$, respectively.  (b) 3D geometry for a two-qubit system with auxiliary
sites, for which the Coulomb interaction between the two charges is constant if 
both are confined to their respective qubit subspace comprised of dots 1 and 2, 
but differs from the Coulomb coupling between the auxiliary dots provided that 
$\sqrt{4b^2+2c^2}\neq 2a$.  Note that the inter-dot distances satisfy
$\norm{\vec{s}_d^{(1)}-\vec{s}_{d'}^{(2)}}=\sqrt{4b^2+2c^2}$ and 
$\norm{\vec{s}_3^{(1)}-\vec{s}_d^{(2)}}=\norm{\vec{s}_d^{(1)}-\vec{s}_3^{(2)}}
=\sqrt{(a+b)^2+c^2}$ for $d,d'=1,2$.}
\label{fig:arch}
\end{figure*}

Rather than choose an arbitrary interaction, it is more useful to consider geometries
where the action of the Coulomb force on the qubit states is trivial.  One way this 
might be achieved is by using shielding to eliminate direct interactions between 
qubits but not between the auxiliary sites as shown in Fig.~\ref{fig:arch} (a). Such
a system could be fabricated in a GaAs 2DEG system~\cite{PRL91n226804,RMP75p001} or a
pillar system~\cite{SCI297p1313}, for instance.  An alternative is to choose a geometry
for which the effect of the Coulomb interaction on the dynamics of the qubit subspace 
is trivial, resulting only in a global phase factor, as in Fig.~\ref{fig:arch} (b). 
To fabricate such a structure, we would propose an extension of the atomic placement 
techniques used for Si:P systems~\cite{PRL91n136104,QIC1p129}.  Both designs have the
advantage of allowing the implementation of controlled two-qubit gates using simple 
pulse sequences.  Note that using auxiliary sites has also been proposed as a way to 
allow generalized readout of quantum information in different bases~\cite{PRB70n041305}.

The Hilbert space for two qubit-plus-auxiliary units is spanned by the states 
$\ket{dd'}$ for $d,d'=1,2,3$, where $\ket{d}_k$ denotes the $d$th basis state 
for the $k$th unit, and $\ket{dd'}=\ket{d}_1\otimes\ket{d'}_2$ are the tensor
product states as usual.  The total Hamiltonian of this system is
\begin{equation} \label{eq:Htot2}
  \op{H} = \op{H}^{(1)} \otimes \op{I}_3 + \op{I}_3 \otimes \op{H}^{(2)} 
           + \op{H}_C,
\end{equation}
where $\op{I}_3$ is the identity matrix in dimension three, $\op{H}^{(k)}$ for
$k=1,2$ is the local Hamiltonian for the $k$th qubit-plus-auxiliary unit specified
in Eq.~(\ref{eq:Hloc1}), and $\op{H}_C$ is the Coulomb Hamiltonian
\begin{equation} \label{eq:HC}
  \op{H}_C = \sum_{d,d'=1}^3  \gamma_{dd'} \ket{dd'}\bra{dd'}.
\end{equation}
The Coulomb interaction energies $\gamma_{dd'}$ are given by 
\begin{equation}
  \gamma_{dd'} = \frac{e^2}{4\pi\eps} 
                 \norm{\vec{s}_d^{(1)}-\vec{s}_{d'}^{(2)}}^{-1},
\end{equation}
where $\vec{s}_d^{(k)}$ denotes the position of the $d$th quantum dot in the 
$k$th qubit-plus-auxiliary unit, $\eps$ is the applicable dielectric constant,
and $e$ is the electron charge.  In pure silicon we have $\eps=11.8\eps_0$,
where $\eps_0$ is the dielectric constant in vacuum.  

If two sites $\vec{s}_d^{(1)}$ and $\vec{s}_{d'}^{(2)}$ are separated by a 
sufficiently thick, grounded metal barrier then $\gamma_{dd'}=0$.  Hence, the 
Coulomb interactions for the shielded 2D geometry in Fig.~\ref{fig:arch} (a) 
effectively vanish except for $\gamma_{33}$.  Similarly, for the 3D geometry 
shown in Fig.~\ref{fig:arch} (b), symmetry implies 
\begin{eqnarray*}
 \gamma_{11}=\gamma_{12}=\gamma_{21}=\gamma_{22} &\equiv& \gamma_1, \\
 \gamma_{13}=\gamma_{23}=\gamma_{31}=\gamma_{32} &\equiv& \gamma_2. 
\end{eqnarray*} 
We can cancel the effect of the Coulomb interaction between qubit states by
applying suitable bias voltages to the energy shift gates to offset the energy 
levels $\eps_1^{(k)}$ and $\eps_2^{(k)}$ by $-\gamma_1/2$ and $\eps_3^{(k)}$ by
$\gamma_1/2-\gamma_2$ for $k=1,2$.  Thus, the Coulomb interaction Hamiltonian 
becomes
\begin{equation}
  \op{H}_C = \gamma_{\rm eff} \ket{33}\bra{33},
\end{equation}
where $\gamma_{\rm eff}$ is the effective Coulomb coupling between the auxiliary
states, i.e., $\op{H}_C$ acts trivially on the system except if both electrons are
in the auxiliary states.

For a 2D architecture with shielding $\gamma_{\rm eff}$ will usually be less than 
the free-space Coulomb interaction $\gamma_{33}$ due to screening and image charges.
For a 3D geometry with energy offsets $\gamma_{\rm eff}$ will be less than or equal
to $\gamma_{33}-2\gamma_2+\gamma_1$, with equality if the screening due to control 
electrodes etc.\ is negligible.  

For convenience, we choose the effective Coulomb energy $\gamma_{\rm eff}$ between
the auxiliary sites as the unit of energy such that the free evolution Hamiltonian
of the two-qubit plus auxiliary system is $\op{H}_0=\ket{33}\bra{33}$.  All tunnelling
rates are given in units of $\gamma_{\rm eff}/\hbar$ and the canonical time unit is 
$\hbar/\gamma_{\rm eff}$.

%~\footnote{%
%The Coulomb energy for two quantum dots in pure silicon 20 nm apart is about 6.1 
%meV, and the corresponding time unit $\gamma_{\rm eff}/\hbar \approx 108$ fs.}.

%%%%%%%%%%%%%%%%%%%%%%%%%%%%%%%%%%%%%%%%%
\section{Controlled Two-Qubit Phase Gate}
\label{sec:phaseGate}
%%%%%%%%%%%%%%%%%%%%%%%%%%%%%%%%%%%%%%%%%

\begin{figure}
\myincludegraphics[width=3.3in]{figures/pdf/energy.pdf}{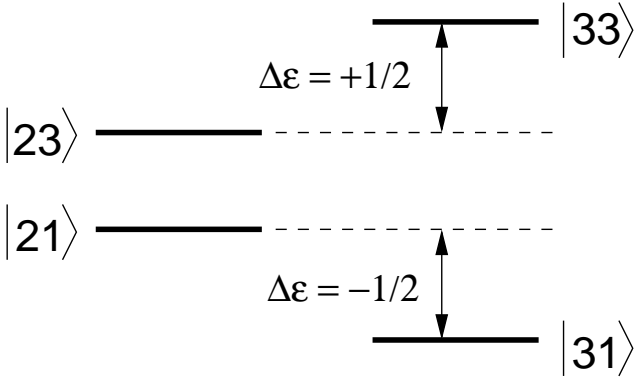}
\caption{Energy level configuration during the second step:  By applying an energy
bias of $\eps_2^{(1)}=1/2$ we ensure that the energy gap between states $\ket{23}$ 
and $\ket{33}$, and $\ket{21}$ and $\ket{31}$ is $\pm 1/2$, respectively.} 
\label{fig:energy}
\end{figure}

To motivate the design of a two-qubit phase gate, we note that the phase acquired
by an electron tunnelling between two quantum dots depends on the energy difference
between them.  In our system the energy differences are determined by a combination
of energy bias voltages and the Coulombic interaction between the auxiliary sites.
Hence, by adjusting the voltage on the energy bias gate $S_2^{(1)}$, for instance, 
we can shift of the energy of state $\ket{2}_1$ such that the energy differences
between states $\ket{21}$ and $\ket{31}$, and $\ket{23}$ and $\ket{33}$, respectively,
are equal in magnitude but opposite in sign, as illustrated in Fig.~\ref{fig:energy}.
In this case the phase acquired by a charge tunnelling between dots~2 and 3 of qubit~1
will have the same magnitude regardless of the population of state $\ket{3}_2$, but
the latter will determine its sign.  This observation is central to the operation of
our controlled phase gate.  

To achieve a maximally entangling gate, the acquired phases in both cases must differ
by an integer multiple of $\pi$.  Finally, except for the acquired phase, the charge 
must return to its initial state $\ket{2}_1$ at the end of the tunnelling process, i.e., 
no population must remain in the auxiliary state $\ket{3}_1$.  Combining all of these 
requirements leads to the following expressions for the tunnelling rate
\begin{equation} \label{eq:mu1}
   \mu_{23}^{(1)} = \frac{1}{4}\sqrt{\left(\frac{2n}{2k-1} \right)^2-1},
\end{equation}
for suitable positive integers $n$ and $k$ satisfying $2n>2k-1$, and tunnelling time
\begin{equation} \label{eq:t2}
   \tau_2 = \frac{4\pi n}{\sqrt{16 \left(\mu_{23}^{(1)} \right)^2 +1}}
          = 2\pi(2k-1).
\end{equation}
A detailed explanation of these results is provided in appendices~\ref{app:swap2} and
\ref{app:step4}.  Furthermore, the phase acquired by state $\ket{21}$ is $[2n-(2k-1)]
\pi/2$, while that of state $\ket{23}$ is $[2n+(2k-1)]\pi/2$, and hence the phase 
difference is $(2k-1)\pi \equiv \pi$ as desired.

In the absence of constraints on the tunnelling rates and pulse lengths, the gate 
operation time is optimized if we choose $n=k=1$ and $\mu_{23}^{(1)}=\sqrt{3}/4$. 
For details about how to choose $n$ and $k$ when there are constraints on the 
tunnelling rates and pulse lengths, see appendix~\ref{app:constraints}.

\begin{figure}
\myincludegraphics[width=3.3in]{figures/pdf/scheme.pdf}{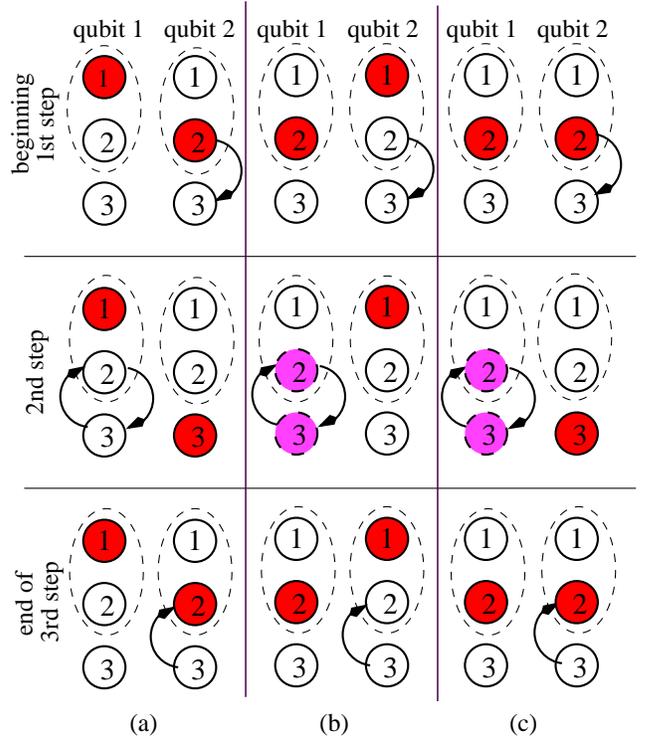}
\caption{Schematic representation of the phase gate operation for three initial
basis states (a) $\ket{12}$, (b) $\ket{21}$, and (c) $\ket{22}$.  The operation 
for $\ket{11}$ has been omitted because it is trivial.  The filled (red) dots 
show the positions of the electrons at the beginning of the first step (top), 
during the second step (middle), and after completion of the third step (bottom). 
For initial configurations $\ket{21}$ (b) and $\ket{22}$ (c), the first charge 
is in a superposition of states $\ket{2}_1$ and $\ket{3}_1$ during the second 
step, and hence acquires a phase conditional on the population of state $\ket{3}_2$, 
which is indicated by by lighter (pink) shading.  Notice, however, that the 
electron starts and ends in state $\ket{2}_1$, except for the conditional phase
acquired.}\label{fig:gate-scheme}
\end{figure}

\begin{figure}
\myincludegraphics[width=3.3in]{figures/pdf/trajct1-ideal.pdf}{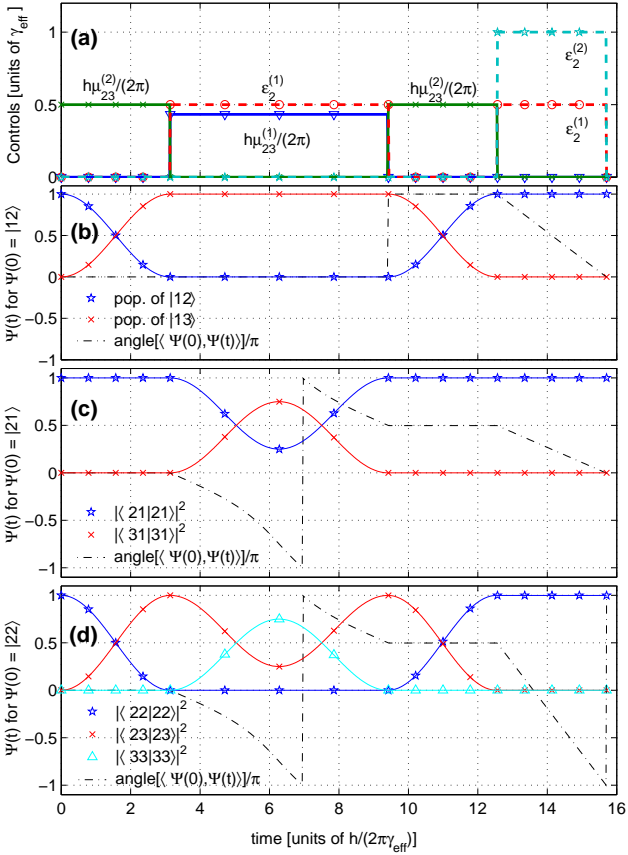}
\caption{Operation of the proposed controlled phase gate for ideal controls with
no constraints.  Shown are the control parameter settings (a) and the corresponding
evolution of the initial states (b) $\ket{12}$, (c) $\ket{21}$ and (d) $\ket{22}$.
The evolution of $\ket{11}$, being trivial, has been omitted.  In each case, only
the relevant populations and acquired phases are plotted.}
\label{fig:gate-ideal}
\end{figure}

More explicitly, a maximally entangling controlled two-qubit phase gate 
\begin{eqnarray}
   \op{U}_{\rm phase} &=&  \ket{11}\bra{11}+\ket{12}\bra{12}
                           +\ket{21}\bra{21}-\ket{22}\bra{22} \nonumber \\
                      &=&  \diag(\op{I}_2,\op{Z}_{12}).
\end{eqnarray}
can be realized as follows: 
\begin{enumerate}
\item Acting on the \emph{second} qubit, swap the populations of the states 
      $\ket{2}_2$ and $\ket{3}_2$ by lowering the tunnelling barrier between 
      the 2nd and 3rd quantum dot for time $\tau_1=\pi/(2\mu_{23}^{(2)})$, 
      where $\mu_{23}^{(2)}$ is the tunnelling rate.

\item Acting on the \emph{first} qubit, simultaneously raise the ground state 
      energy of the 2nd dot by $\eps_2^{(1)}=1/2$ and lower the tunnelling 
      barrier between the 2nd and 3rd dot to achieve the tunnelling rate 
      given by Eq.~(\ref{eq:mu1}) for the time specified by Eq~(\ref{eq:t2}).

\item Acting on the \emph{second} qubit, repeat the first step to swap the
      populations of states $\ket{2}_2$ and $\ket{3}_2$ back.

\item Acting simultaneously on both qubits, shift the energy of states 
      $\ket{2}_1$ and $\ket{2}_2$ by 
      $\eps_2^{(1)}=\ell\pi/\tau_4$ and 
      $\eps_2^{(2)}=\pi/\tau_4$, respectively, for some time $\tau_4$, where 
      $\ell=1/2-(n+k)\mbox{ mod }2$.
      (See appendix~\ref{app:step4}.)
\end{enumerate}
The first three steps are illustrated in Fig.~\ref{fig:gate-scheme}, and 
Fig.~\ref{fig:gate-ideal} shows results for a simulation of the gate for ideal 
pulses and no constraints.

Since this gate combined with local unitary operations as described in 
Sec.~\ref{sec:local} is universal, we can implement any desired two-qubit gate.  
For instance, a controlled-NOT gate is performed simply by applying a Hadamard
transformation on the second qubit before and after the pulse sequence above.

\begin{figure}
\hspace*{-0.3in}
\myincludegraphics[width=3.5in]{figures/pdf/bloch1-ideal.pdf}{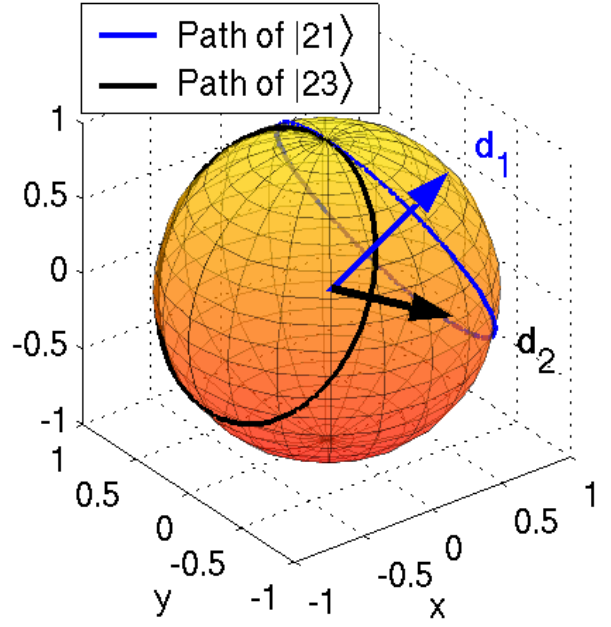}
\caption{Trajectory of the Bloch vectors associated with the two-level subsystems
$S_1$ and $S_2$ for ideal phase gate.  For $n=k=1$ the Bloch vectors $\vec{s}_1(t)$ 
(cyan) and $s_2(t)$ (black) rotate about $\vec{d}_1$ and $\vec{d}_2$, respectively, 
simultaneously completing a single closed loop on the surface of the Bloch sphere
and sweeping out solid angles of $\pi$ (area ``inside'' the blue loop) and $3\pi$ 
(area ``outside'' the black loop), respectively.  Hence, the difference between the 
areas is $2\pi$, and the conditional phase acquired by state $\ket{2}_1$ depending on 
whether state $\ket{3}_2$ is occupied will differ by $\pi$.} \label{fig:gate-bloch}
\end{figure}

Further analysis shows that the conditional phase acquired by state $\ket{2}_1$
is exactly twice the geometric phase acquired by the two-level subsystem $S_1=
\{\ket{21},\ket{31}\}$ and $S_2=\{\ket{23},\ket{33}\}$, respectively.  To see 
this, note that the Hamiltonians for $S_1$ and $S_2$ are 
\begin{eqnarray}
  \op{H}_{S_1}  & = & \frac{1}{4} \op{I}_2 
                    + \mu \op{\sigma}_x + \frac{1}{4}\op{\sigma}_z \\
  \op{H}_{S_2} & = & \frac{3}{4} \op{I}_2 
                    + \mu \op{\sigma}_x - \frac{1}{4}\op{\sigma}_z 
\end{eqnarray}
with $\mu=(\sqrt{(2n)^2/(2k-1)^2-1})/4$, and we can visualize their evolution on 
the Bloch sphere as in Fig.~\ref{fig:gate-bloch}.  Let $\vec{s}=(s_x,s_y,s_z)$ 
with $s_\alpha=\Tr(\sigma_\alpha\rho)$ for $\alpha\in\{x,y,z\}$ be the usual Bloch vector.  
The initial states $\ket{21}$ and $\ket{23}$ for $S_1$ and $S_2$, respectively, 
correspond to the Bloch vector $\vec{s}_0=(0,0,1)$, and their evolution in $\RR^3$ 
to a rotation about the axes $\vec{d}_1=2(\mu,0,1/4)$ and $\vec{d}_2=2(\mu,0,-1/4)$, 
respectively.  

The pure-state non-adiabatic, cyclic geometric phase~\cite{PRL58p1593} is 
$\phi_{\rm geom}=\frac{\Omega}{2}=2\pi n(1-\cos\theta)$, where $\Omega$ is the
solid angle subtended by the Bloch vector, $n$ is the number of times the Bloch
vector rotates around the axis $\vec{d}$, and $\theta$ is the angle between the 
initial state vector $\vec{s}_0$ and the rotation axis $\vec{d}$.  
Both states rotate with the same frequency $\norm{\vec{d}}=\norm{\vec{d'}}=
(\sqrt{1+16\mu^2})/2=n/(2k-1)$, and hence the relative geometric phase depends 
on $\theta$.  Noting that
\begin{equation}
  \cos(\theta_1) = \frac{\vec{d}_1\cdot\vec{s}_0}{\norm{\vec{d}_1}}=\frac{2k-1}{2n}
              = -\cos(\theta_2),
\end{equation}
we see that the respective geometric phases acquired by $S_1$ and $S_2$ are $2\pi n
[1-(2k-1)/(2n)]=\pi[2n-(2k-1)]$ and $2\pi n[1+(2k-1)/(2n)]=\pi[2n+(2k-1)]$, i.e., 
exactly twice the conditional phases acquired by states $\ket{21}$ and $\ket{23}$.

%%%%%%%%%%%%%%%%%%%%%%%%%%%%%%%%%%%%%%%%%%%%%%%%%%%%%%%%%%%%%%%%
\section{Realistic Controls and Correction of Systematic Errors}
\label{sec:errors}
%%%%%%%%%%%%%%%%%%%%%%%%%%%%%%%%%%%%%%%%%%%%%%%%%%%%%%%%%%%%%%%%

\begin{figure}
\myincludegraphics[width=3.3in]{figures/pdf/SWP.pdf}{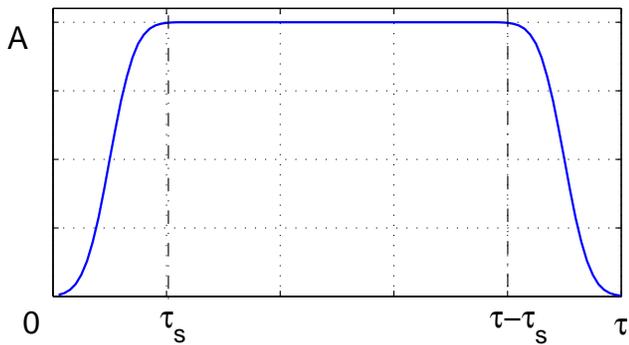}
\caption{Realistic square-wave pulse of amplitude $A$ and length $\tau$ with rise 
and decay time $\tau_s$, modelled using error functions.} 
\label{fig:SWP}
\end{figure}

\begin{figure*}
\begin{tabular}{cc}
(I) & (II) \\
\myincludegraphics[width=3.3in]{figures/pdf/trajct1-error.pdf}{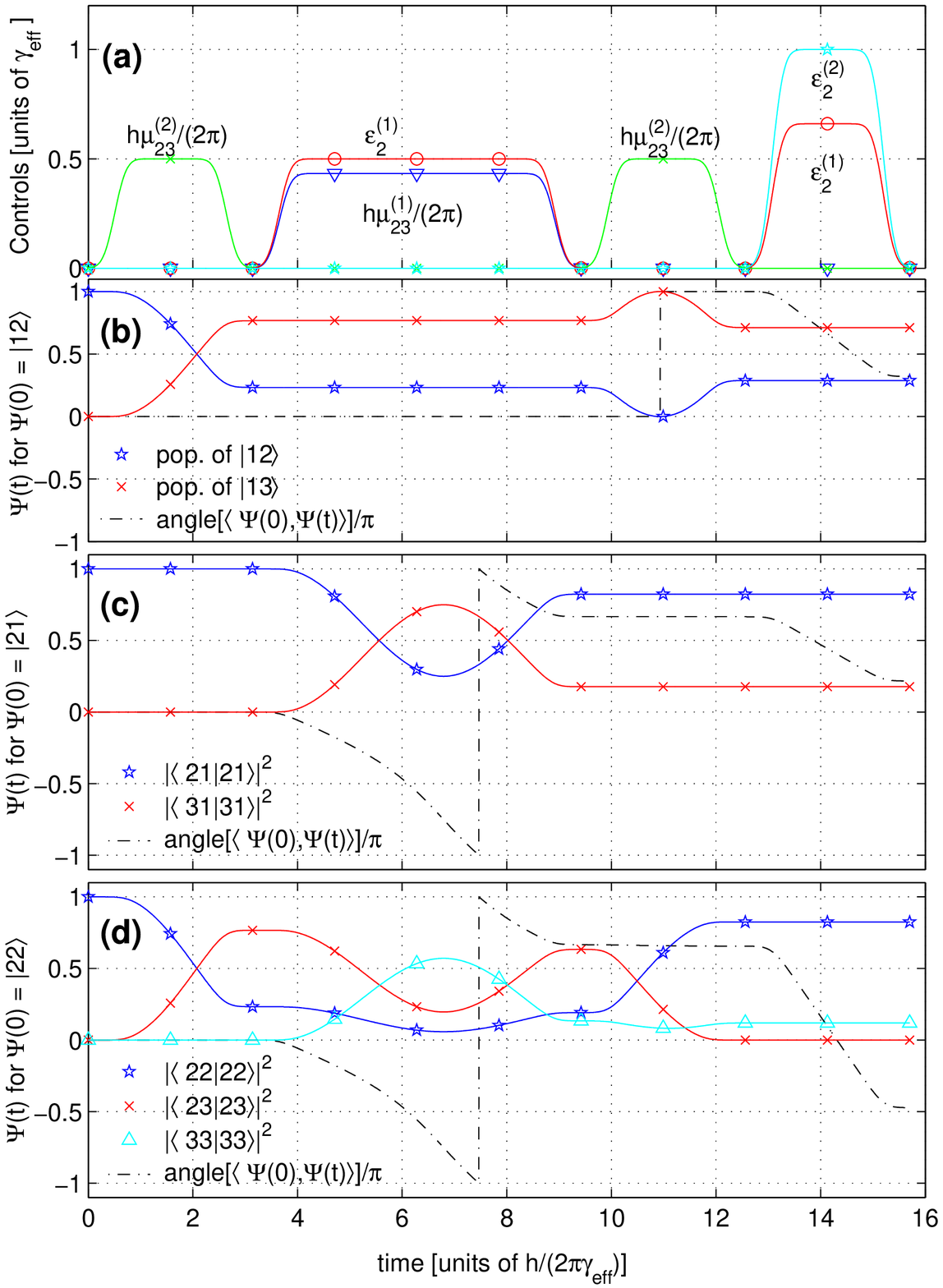} &
\myincludegraphics[width=3.3in]{figures/pdf/trajct1-correct.pdf}{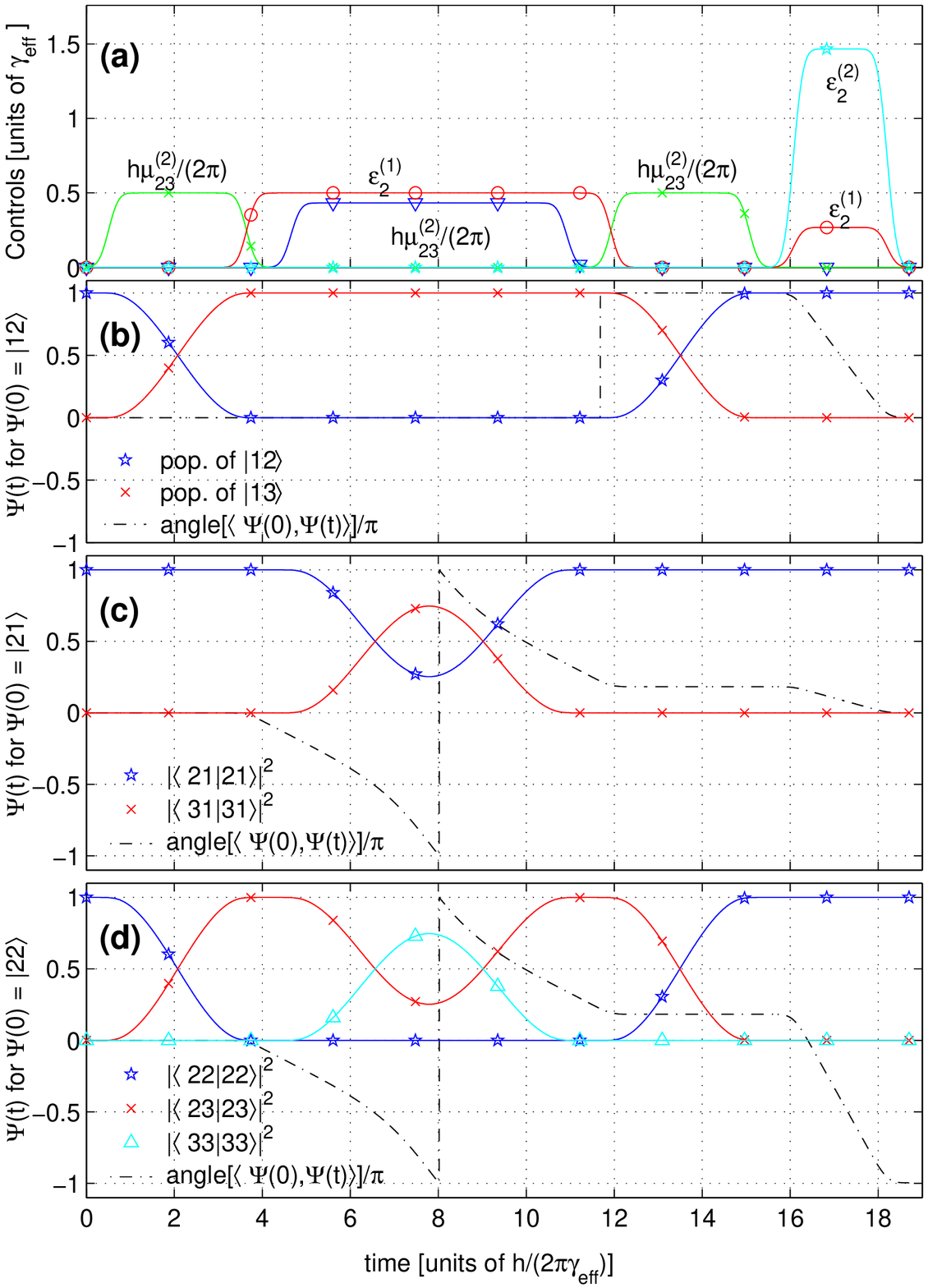} \\
{\sf\bf (e)} 
\myincludegraphics[width=3.1in]{figures/pdf/bloch1-error.pdf}{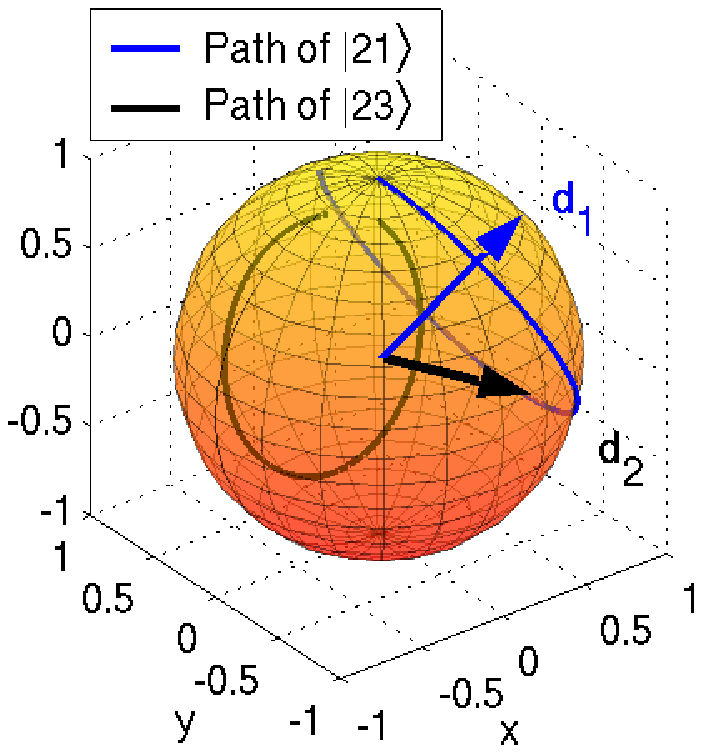} &
{\sf\bf (e) } 
\myincludegraphics[width=3.1in]{figures/pdf/bloch1-correct.pdf}{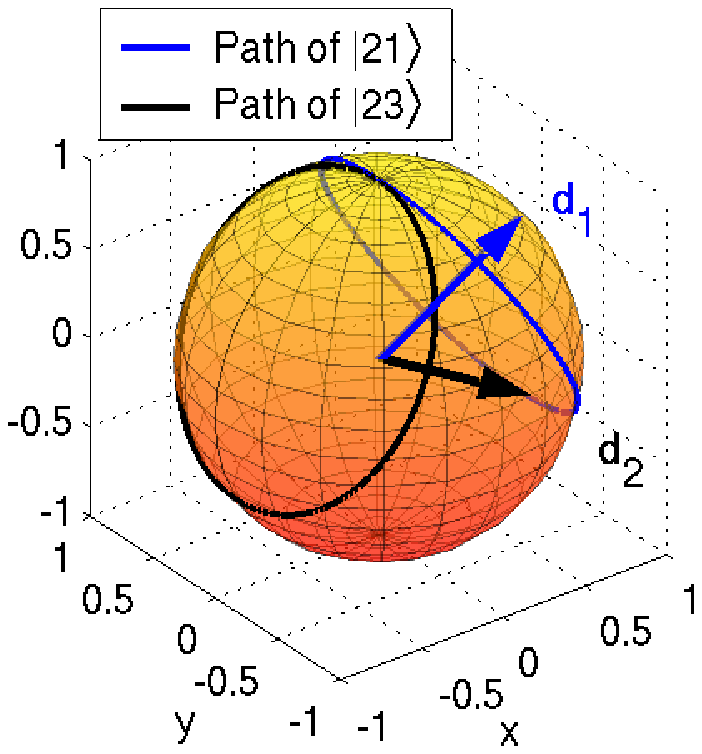}
\end{tabular}
\caption{Control parameter settings (a) and evolution of the initial states (b) 
$\ket{12}$, (c) $\ket{21}$ and (d) $\ket{22}$, as well as (e) evolution of the 
two-level subsystems $S_1$ and $S_2$ on the Bloch sphere under the operation of
the controlled phase gate ($n=k=1$) for controls with rise and decay time $\tau_s=1$
$[\hbar/\gamma_{\rm eff}]$ without correction of systematic errors (I) and with 
corrections described (II).  Due to the rise and decay time of the controls, the
trajectories of the Bloch vectors in (I) do not form closed loops, and the system
does not return to its initial state (modulo the desired phase change).  However, 
when the rise and decay times of the pulses are taken into account, population 
loss to the auxiliary states is negligible and the trajectories of the Bloch 
vectors again form closed loops (II).} 
\label{fig:gate-error}
\end{figure*}

So far we have assumed ideal, piecewise-constant control Hamiltonians.  In reality,
however, the pulses will have finite rise and fall times, for example, as shown in 
Fig.~\ref{fig:SWP}.  A na\"{i}ve implementation of the above scheme would perforce 
introduce systematic errors.  These errors are evident in Fig.~\ref{fig:gate-error} 
(I), which show the evolution of the system for the same parameter values as in 
Fig.~\ref{fig:gate-ideal} except for pulse rise and decay times of $\tau_s=1$ time 
unit.  Comparing the trajectories shows that there are significant population and 
phase errors.  In particular, the populations of the states $\ket{13}$, $\ket{31}$ 
and $\ket{33}$ do not return to $0$, i.e., population is lost to the auxiliary states,
a potentially difficult error to correct.

To quantitatively compare the gates implemented to the ideal gate we consider the 
\emph{average gate fidelity}~\cite{PRL78p0390,PLA294p0258}
\begin{equation}
  {\cal F} = \overline{\bra{\Psi_{\rm in}}\op{U}^\dagger\op{\rho}_{\rm out}\op{U}
               \ket{\Psi_{\rm in}}}, 
\end{equation}
which is a measure of the overlap of the final state $\op{\rho}_{\rm out}$ with the 
desired target state $\op{U}\ket{\Psi_{\rm in}}$ averaged over all input states. 
Since $\op{\rho}_{\rm out}$ may extend to the auxiliary quantum dots, the fidelity
includes the effect of population losses to these states.

In the example shown in Fig.~\ref{fig:gate-error} pulse rise and decay times of one
time unit reduce the average gate fidelity from one in the ideal case to $\sim 73$~\%. 
Table~\ref{table:trise} shows that even relatively small rise and decay times of the
fields tend to reduce the average gate fidelity noticeably.  This increase in the 
gate error appears to be mainly due to population loss to the auxiliary states.

\begin{table}
\[\begin{array}{|r|*{6}{c|}} \hline
 \tau_s & 0.25   & 0.50   & 0.75   & 1.00   & 1.25   & 1.50      \\\hline
 \E_u & 0.0223 & 0.0825 & 0.1691 & 0.2659 & 0.3559 & 0.4251    \\\hline
 \E_c \times 10^4 & 0.2225 & 0.2249 & 0.2273 & 0.2294 & 0.2311 & 0.2325 \\\hline
\end{array}\]
\caption{Gate error $\E=1-\F$ as a function of the pulse rise and decay time 
$\tau_s$ for non-ideal, uncorrected ($\E_u$) and corrected pulses ($\E_c$) for
simulations with time step $\Delta t=0.005$.} \label{table:trise}
\end{table}

The first step toward improving the results is to realize that rise and decay times 
reduce the total pulse area.  An ideal square-wave pulse ($\tau_s=0$) of duration 
$\tau$ with amplitude $A$ has a pulse area of $A\tau$, while that of a similar pulse
with rise and decay time $\tau_s<\tau/2$ is only $A(\tau-\tau_s)$~\cite{JPA35p8315}.  
Since the pulse area is an important control parameter, we must therefore adjust 
either the field strengths or the pulse lengths, or possibly both.  For the 1st, 
3rd and 4th step of the CPHASE gate correcting the pulse area is all that is needed 
to eliminate systematic errors due to non-zero pulse rise and decay times, and we 
can achieve this either by increasing the field strength or the pulse durations.  

During the crucial 2nd step, the energy of state $\ket{2}_1$ should be kept constant
$\eps_2^{(1)}=1/2$ while tunnelling is enabled, and the value of the tunnelling rate 
$\mu_{23}^{(1)}$ should be close to one of the desired values.  Increasing the field
strengths to compensate for rise and decay times is thus not an option for this step.  
For simplicity, we shall therefore correct the pulse areas by increasing the duration
of each pulse by $\tau_s$, i.e., setting $\tau_k'=\tau_k+\tau_s$, where $\tau_k$ is 
the duration of the corresponding ideal pulse.

To improve the accuracy of the 2nd step further, we raise the energy $\eps_2^{(1)}$ 
at least $\tau_s$ time units before we enable tunnelling between states $\ket{2}_1$ 
and $\ket{3}_1$, and lower it only after the tunnelling barrier has been raised again
to inhibit tunnelling.  Since changing the energy of state $\ket{2}_1$ is a local 
operation on the first qubit, it commutes with the swap operation on states $\ket{2}_2$
and $\ket{3}_2$.  We can therefore begin to raise the energy of state $\ket{2}_1$ 
before the 1st swap operation is completed, and we can begin the 3rd swap operation
before $\eps_2^{(1)}(t)$ has reached zero.  However, the increased duration of the 
energy shift will induce an additional local phase shift, which has to be taken into 
account in the final step to achieve the desired gate.  Let $t_1=\tau_1'-\tau_s$ and 
$t_2=\tau_1'+\tau_2'+\tau_s$ and
\begin{equation}
  \Delta\phi = \int_{t_1}^{t_2} \!\!\!\eps_2^{(1)}(t)\,dt - \frac{1}{2}\tau_2,
\end{equation} 
where $\tau_2$ is the time required to complete the second swap operation for ideal, 
piecewise constant control pulses.  Then the local phase rotation on the first qubit
required in the final step is $\Phi=\ell\pi-\Delta\phi$ instead of $\ell\pi$, where 
$\ell=\frac{1}{2}-(n+k)\mbox{ mod }2$ as before.  Thus, in the final step the energy
of state $\ket{2}_1$ must be raised by 
\begin{equation}
    \eps_2^{(1)} = \frac{\pi\ell-\Delta\phi}{\tau_4}
\end{equation}
where $\tau_4$ is the duration of the step for ideal, piecewise constant controls.

Numerical simulations indicate that these corrections greatly enhance the performance
of the phase gate.  For instance, Table~\ref{table:trise} shows that the average gate
error $\E_c$ with these corrections is less than $10^{-4}$ for \emph{all} rise and
decay times $\tau_s$~\footnote{These figures are for numerical simulations with time
step $\Delta t=0.005$ and can be improved further by decreasing $\Delta t$.} and 
Fig.~\ref{fig:gate-error} (II) shows that the trajectories very closely match those 
of the ideal gate.

%%%%%%%%%%%%%%%%%%%%%%%%%%%%%%%%%%%%%%%%%%%%%%%%%%%%%%%
\section{Gate Operation Times and Physical Constraints} 
\label{sec:constraints}
%%%%%%%%%%%%%%%%%%%%%%%%%%%%%%%%%%%%%%%%%%%%%%%%%%%%%%%

In the hypothetical case of no constraints on the tunnelling rates, energy shifts 
and pulse lengths considered so far, the first and third swap operation as well as
the local phase rotations could, in principle, be implemented arbitrarily fast and
the gate operation time would be limited mainly by the time required to complete 
the second step, i.e., $\tau\approx \tau_2=2\pi$ time units.  In reality, however, 
the gate operation time is usually limited by physical and technical constraints 
that impose, for instance, a minimum pulse length $\tau_{\rm min}$ and maximum 
tunnelling rate $\mu_{\rm max}$ between the qubit and auxiliary quantum dot.  

To explore the consequences of finite tunnelling rates and switching speeds, let 
us consider a specific example of a 2D charge qubit architecture as shown in 
Fig.~\ref{fig:arch} (a) with auxiliary quantum dots spaced about $2a=170$~nm apart%
~\footnote{This value seems realistic to allow for a metal barrier sufficiently thick
to provide adequate screening and sufficient distance between the quantum dots and 
the barrier to minimize barrier-induced decoherence.}.  
Neglecting screening effects, the Coulomb energy in this case is $\gamma_{\rm eff}
\approx 0.718$~meV, and the characteristic time scale $\hbar\gamma_{\rm eff}^{-1}
\approx 1$~ps~\footnote{In practice, the effective Coulomb interaction should be
determined experimentally for a given system.}  Thus, theoretically, two-qubit gate
operation times of less then 20~ps could be achieved, as the previous sections show,
and if one assumes that local operations can be realized arbitrarily fast then gate
operation times of less than 7~ps would be possible.

\begin{figure*}
\begin{tabular}{cc}
{\sf\bf(I)} & {\sf\bf(II)} \\
\myincludegraphics[width=3.3in]{figures/pdf/trajct2a.pdf}{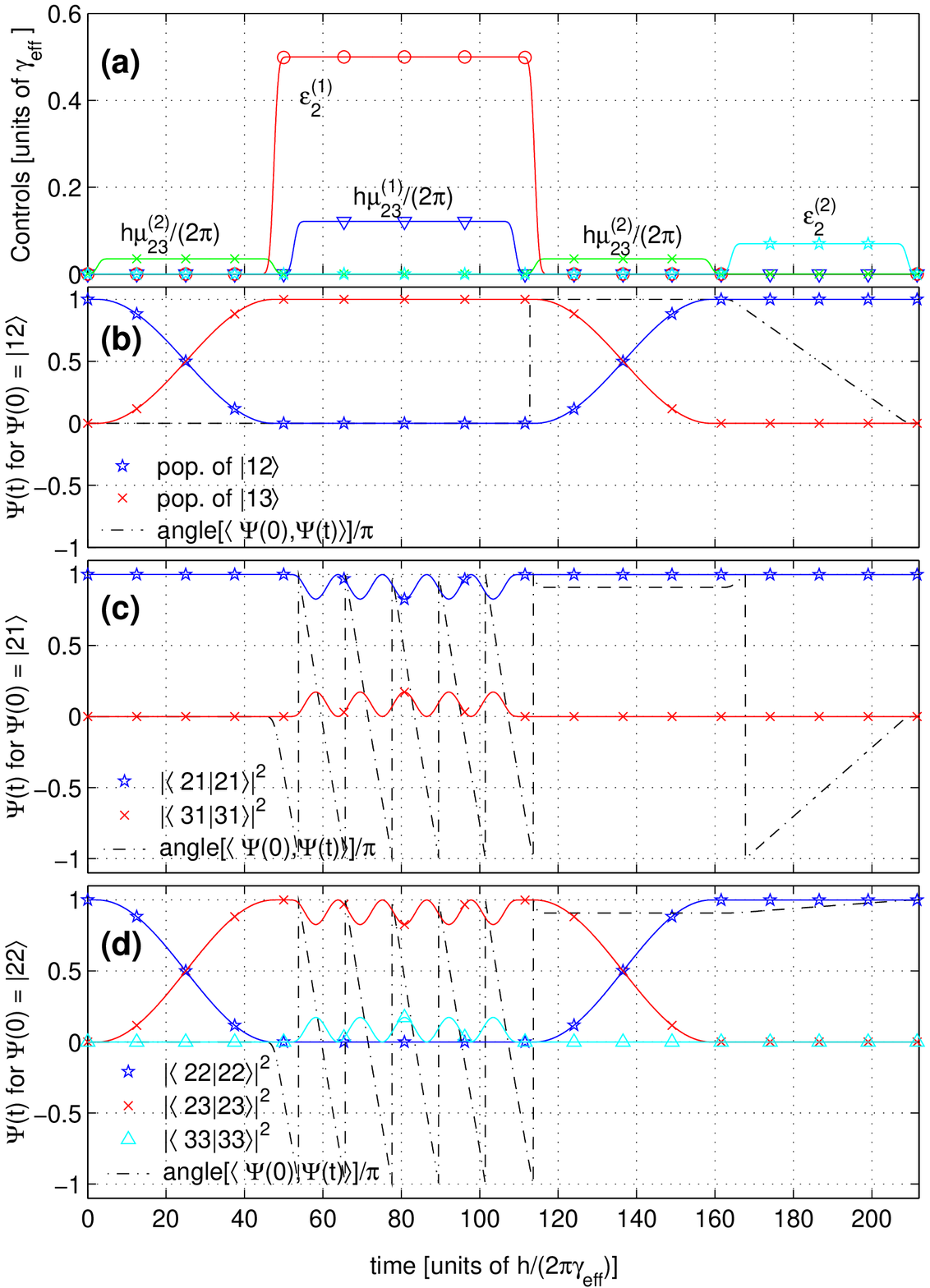}&
\myincludegraphics[width=3.3in]{figures/pdf/trajct2b.pdf}{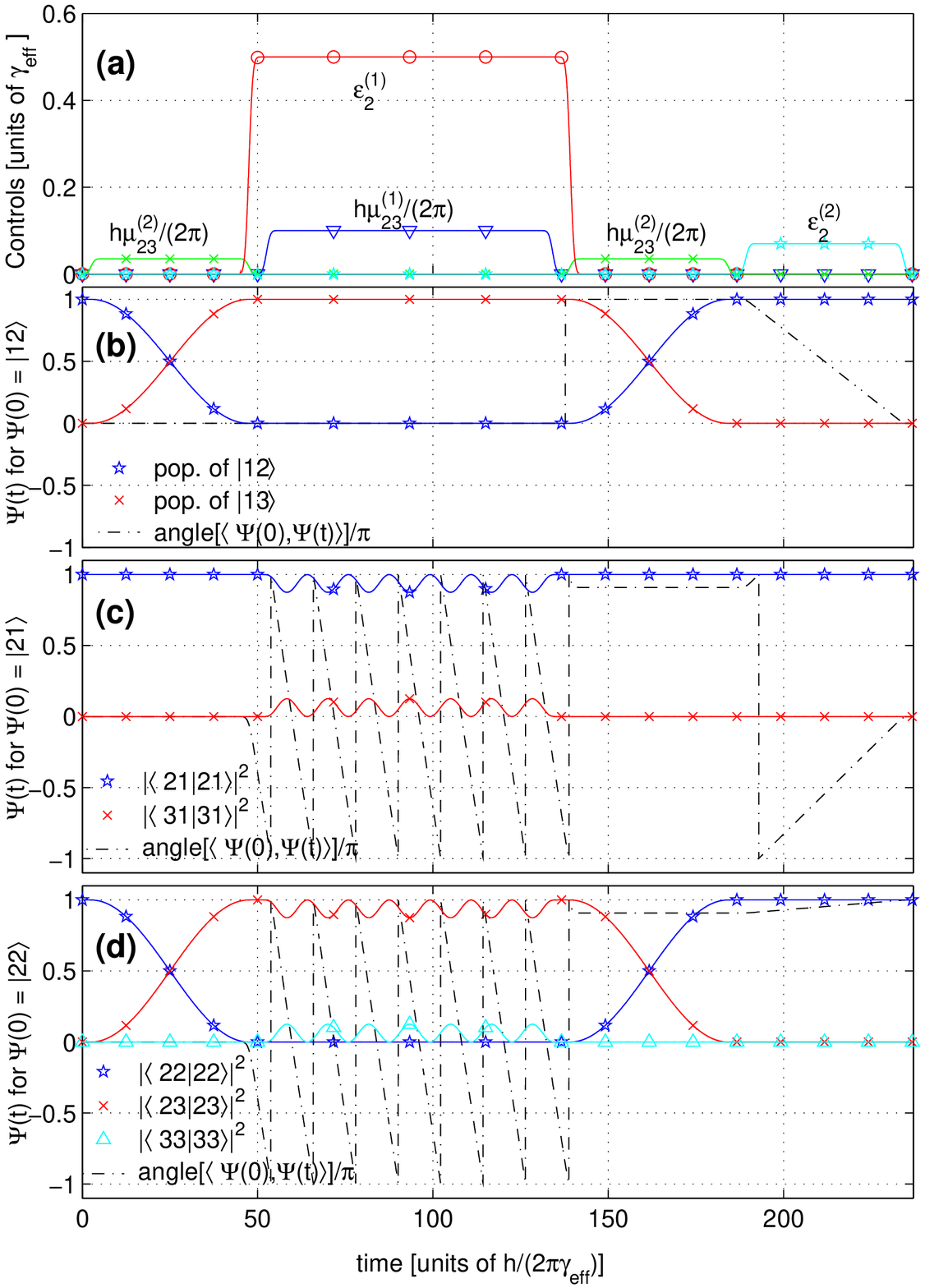}\\
{\sf\bf(e)}\myincludegraphics[width=3.1in]{figures/pdf/bloch2a.pdf}{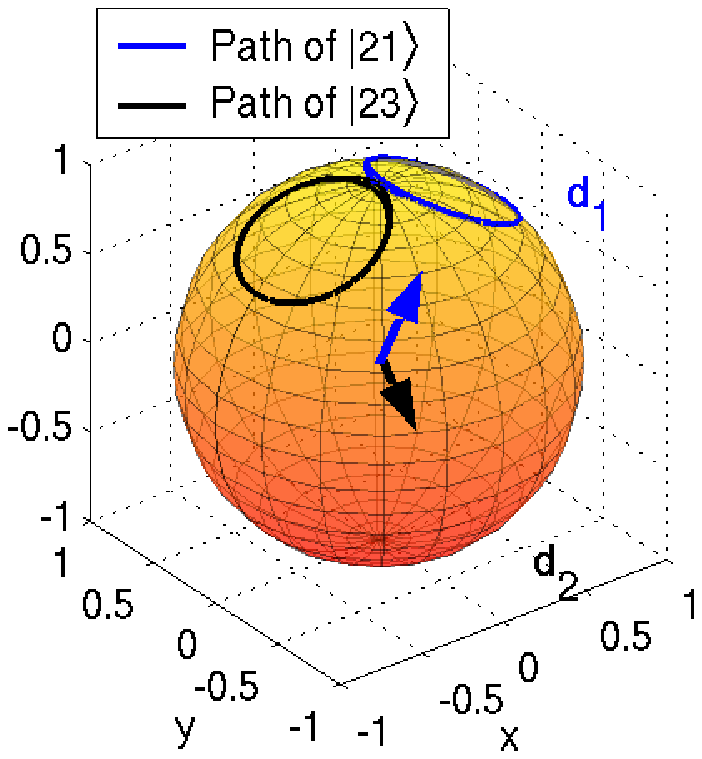}&
{\sf\bf(e)}\myincludegraphics[width=3.1in]{figures/pdf/bloch2b.pdf}{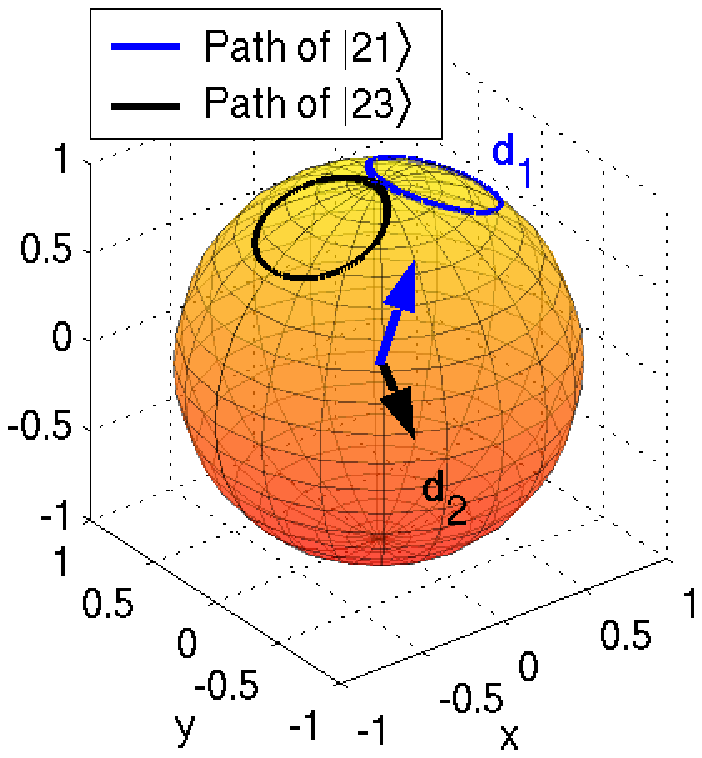}
\end{tabular}
\caption{Control settings (a) and corresponding evolution of the initial states 
(b) $\ket{12}$, (c) $\ket{21}$, and (d) $\ket{22}$, as well as evolution of the 
two-level subsystems $S$ and $S'$ on the Bloch sphere (e) under the operation of 
the controlled phase gate when (I) minimum pulse length constraints of $\tau_{\rm min}
\ge 50$~ps necessitate $n=k=5$, and (II) simultaneous pulse length and tunnelling
rate constraints $\tau_{\rm min}\ge 50$~ps and $\mu_{23}^{(1)}\le 10^{-11}$~rad~s$^{-1}$ 
necessitate $n=k=7$.  Note the decrease in the angle $\theta_1$ (increase in 
$\theta_2$) between the rotation axis $\vec{d}_1$ ($\vec{d}_2$) and the (positive) 
$z$-axis compared to the $n=k=1$ case, and the resulting greater difference in 
the areas swept out by the Bloch vectors $\vec{s}_1$ (blue) and $\vec{s}_2$ (black)
in a single loop: $\pi/5$ versus $19\pi/5$ in (I); $\pi/7$ versus $27\pi/7$ in 
(II).  Consequently, five and seven loops, respectively, are necessary to archive
an area difference that is a multiple of $2\pi$, and the time required to implement
the phase shift increases to $18\pi$ and $26\pi$, respectively.  Also note the 
slight distortion of the loops near the north pole due to the dynamic change of 
the rotation axis during rise and decay periods of the pulses.  Although present 
for all choices of $n$ and $k$, it is more pronounced for larger $n$ and $k$.} 
\label{fig:gate2}
\end{figure*}

However, if the pulse lengths must be $\gtrsim 50$~ps --- about the limit of what 
is accessible with current technology --- then Eq.~(\ref{eq:kn}) shows that the best 
possible choice of the parameters $n$ and $k$ that satisfies this constraint while
minimizing the gate operation time is $k=n=5$.  The total gate operation time in 
this case increases to $3\times 50+18\pi\approx 206.55$~ps --- assuming we can
achieve the required tunnelling rate of $\mu_{23}^{(1)}=(1/4)\sqrt{(10/9)^2-1}
[\gamma_{\rm eff}/\hbar]\approx 1.21\times 10^{11}$~rad~s$^{-1}$.  The evolution
of the system for the resulting gate is shown in Fig.~\ref{fig:gate2} (I).

If the maximum tunnelling rate that can be achieved (e.g., without applying
control voltages that might lead to a breakdown of the oxide layer separating the
silicon substrate and the surface control electrodes, or pulses that might result
in population loss to higher-lying excited states) are lower than this, then the
gate operation time will be increased further.  For example, if we must satisfy 
$\mu_{23}^{(1)}\le 10^{-11}$~rad~s$^{-1}$ then Eqs.~(\ref{eq:k2}), (\ref{eq:n2}) 
show that the best choice for $n$ and $k$ is $n=k=7$, which yields $\mu_{23}^{(1)}
\approx 0.99 \times 10^{11}$~rad~s$^{-1}$, and the gate operation time increases 
to $231.68$~ps, as shown in Fig.~\ref{fig:gate2} (II).

%%%%%%%%%%%%%%%%%%%%%%%%%%%%%%%%%%%%%%%%%%%%
\section{Gate Robustness for Noisy Controls}
\label{sec:noise}
%%%%%%%%%%%%%%%%%%%%%%%%%%%%%%%%%%%%%%%%%%%%

\begin{figure}
\myincludegraphics[width=3.3in]{figures/pdf/fid-noise.pdf}{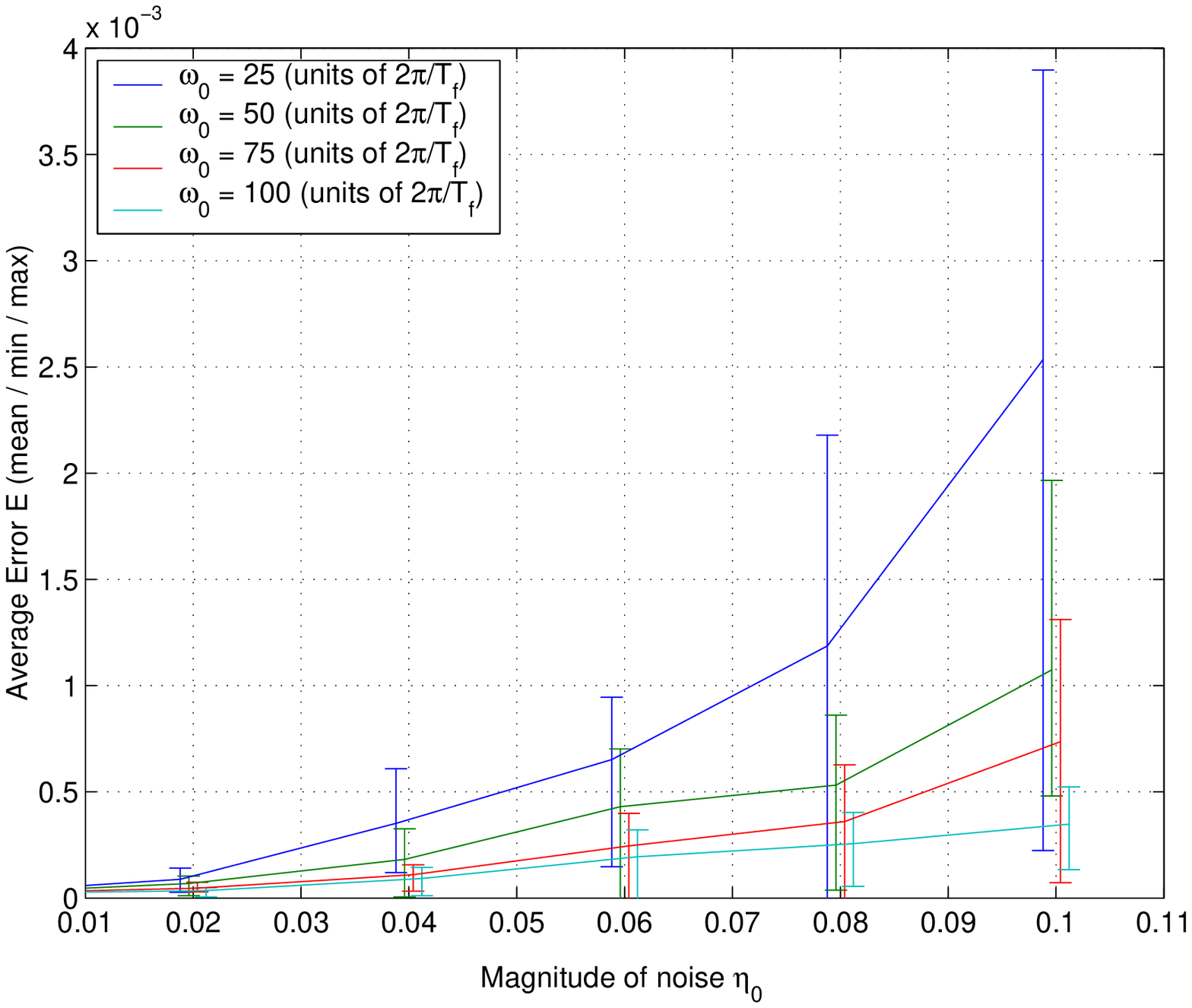}
\caption{Average gate error $\E$ as a function of the magnitude $\eta_0$ of the 
noise for various threshold frequencies $\omega_0$.  The solid curves show the mean 
of $\E$and the error bars indicate the range of $\E$ over 10 simulations.  The data 
were plotted horizontally offset to improve visual clarity.}
\label{fig:noise}
\end{figure}

Another important issue in practice is the robustness of the gate for noisy controls.
To study this effect we artificially added noise to our controls.  Fig.~\ref{fig:noise} 
shows the increase of the gate error $\E$ as a function of the magnitude of the noise
added.  The simulations were performed with a fixed time step of $\Delta t=0.005$, for 
which the average gate fidelity in the absence of noise exceeded $0.9999$.  The noise 
functions $\eta(t)$ were bounded
\begin{equation} 
   |\eta(t)| \le \eta_0
\end{equation} 
with Fourier transforms $\tilde{\eta}(\omega)$ satisfying
\begin{equation}
   |\tilde{\eta}(\omega)| \le \left\{ \begin{array}{lcl} 
   0          && \omega=0 \\
   K/\omega_0 && \omega\le\omega_0 \\
   K/\omega   && \omega>\omega_0.
   \end{array} \right. 
\end{equation}
where $K$ is a constant.  This type of noise corresponds to $1/f$ noise with a
low-frequency cut-off, which is common in electronic devices.

\begin{figure}
\myincludegraphics[width=3.3in]{figures/pdf/noise-trajct.pdf}{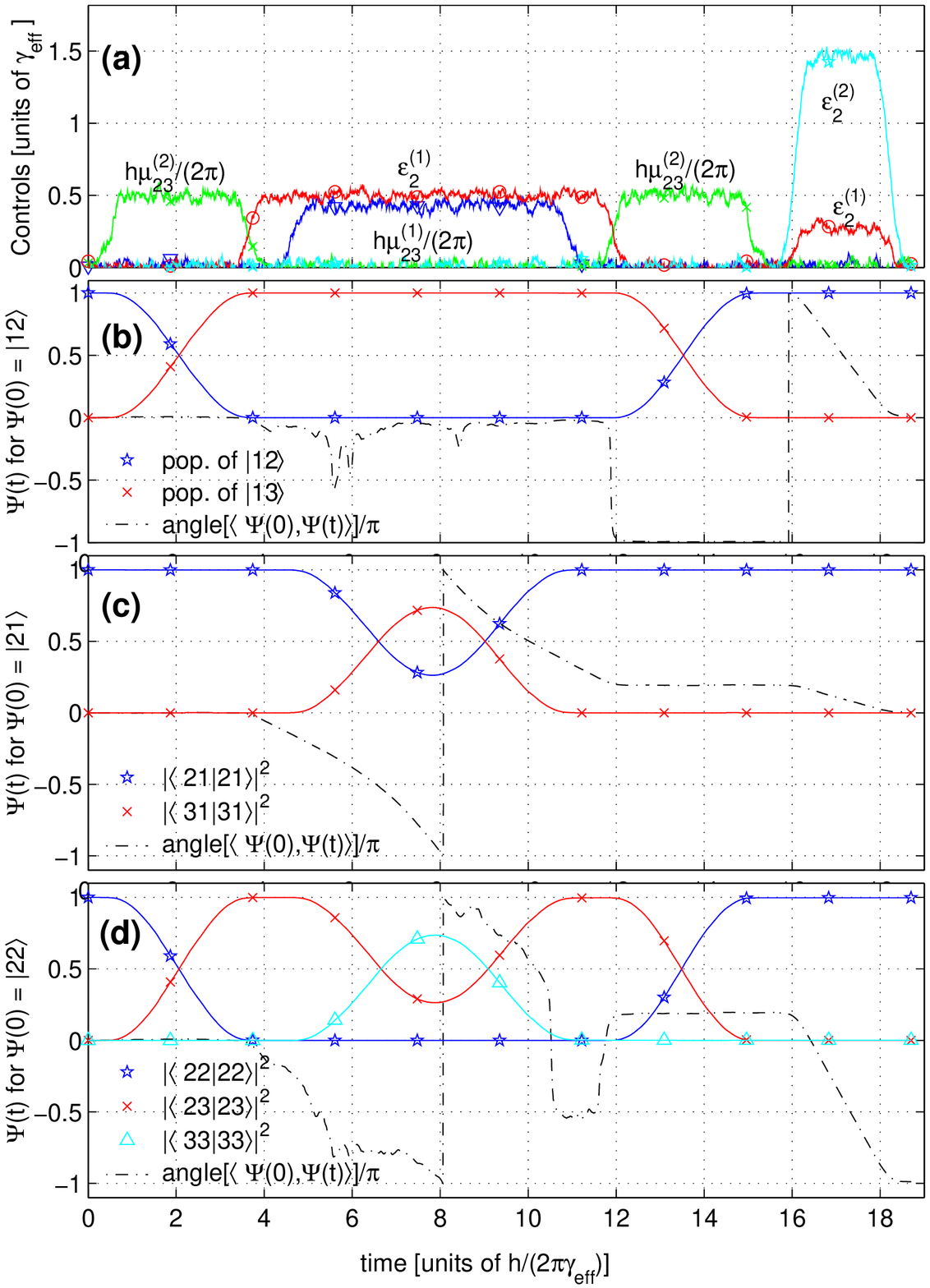}
{\sf\bf (e)}
\myincludegraphics[width=3.15in]{figures/pdf/noise-bloch.pdf}{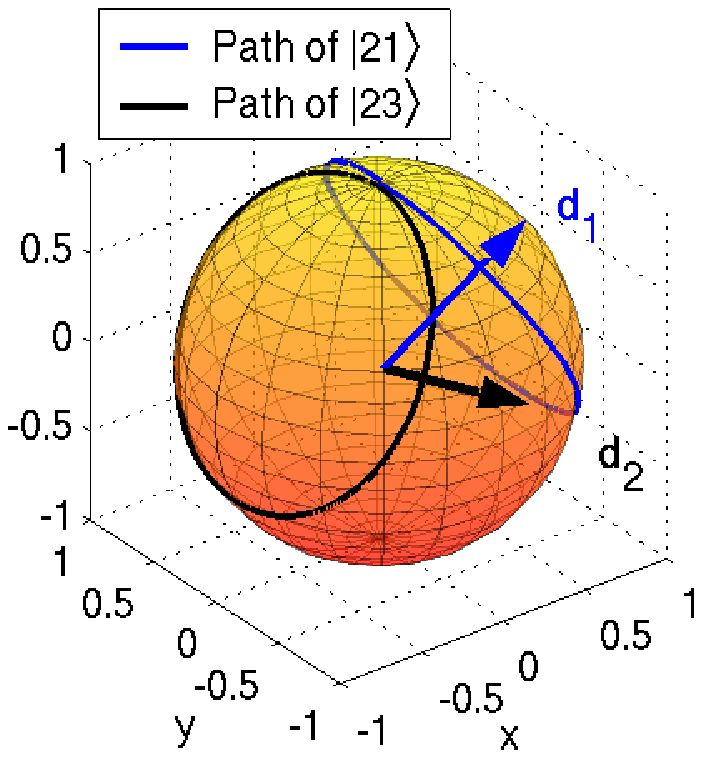}
\caption{Control settings (a) and corresponding evolution of the initial states (b) 
$\ket{12}$, (c) $\ket{21}$ and (d) $\ket{22}$, as well as evolution of the two-level 
subsystems $S_1$ and $S_2$ on the Bloch sphere (e) under the operation of our phase 
gate for noisy controls ($\eta_0=0.1$, $\omega_0=50$).  The noise has no noticable 
effect on the evolution of the populations.  The relative phases exhibit some wiggles
but they mostly seem to average out over the duration of the gate.  The average gate 
error was $\E=0.708 \times 10^{-3}$.}\label{fig:gate-noise}
\end{figure}

Although the addition of noise necessarily increase the gate error, the gate appears to
be quite robust to this noise, as a typical example of the evolution of the system for 
noisy controls (Fig.~\ref{fig:gate-noise}) shows.  Not unsurprisingly, the curves for 
varying threshold frequencies $\omega_0$ in Fig.~\ref{fig:noise} suggest that the gate 
is more sensitive to low-frequency than high-frequency noise, as the high-frequency 
components tend to cancel.

%%%%%%%%%%%%%%%%%%%%%%%%%%%%%%%%%%%%%%%%%%%%%%%%%%
\section{Gate Robustness for Imperfect Geometries}
\label{sec:perturb}
%%%%%%%%%%%%%%%%%%%%%%%%%%%%%%%%%%%%%%%%%%%%%%%%%%

A final issue that must be considered is the effect of manufacturing tolerances,
which result in deviations of the qubit register from an ideal specification.
In solid-state charge-based architectures of the types considered herein, the 
main issues appear to be imperfections in the placement or geometry of the quantum
dots.  The former is believed to be especially pronounced for quantum dots based 
on donor impurities in a Silicon matrix, while the latter will be relevant, e.g.,
for manufactured Ga/GaAs heterostructure quantum dots.  As modelling all the 
possible effects of imperfections in the fabrication process for heterostructure 
quantum dots would far exceed the scope of this paper, we shall focus on modelling
the effect of imperfect placement of the quantum dots on the gate performance.

For solid-state architectures based on donor impurities (e.g., Phosphorus) in a
Silicon matrix the accuracy of placement of the donors is limited.  In a shielded
2D architecture inaccurate placement of the donors will mainly cause variations in
the tunnelling rates as the structure of the silicon lattice introduces spatial
oscillations in the donor wavefunctions, and hence tunnelling rates~\cite{CW-comm}
analogous to those seen in exchange systems~\cite{PRL88n027903,PRB68n195209}.  
There will also be minor changes in the strength of the Coulomb interaction between 
the auxiliary sites.  However, the actual Coulomb coupling strengths and tunnelling 
rates as a function of the control voltages applied can, in principle, be determined
using Hamiltonian identification techniques similar to those described in 
Ref.~\cite{PRA69n050306}), for instance, and the control scheme can then easily
be adapted to achieve the desired gate for the actual system. 

The situation is different for 3D architectures [cf.~Fig.~\ref{fig:arch} (b)] that
rely on symmetry to effectively cancel the Coulomb interactions between the qubit 
sites.  Manufacturing tolerances in this case will result in asymmetries of the 
geometry, which cannot be compensated easily, even if they could be identified 
precisely, and perforce reduce the gate fidelity.  To estimate the effect of such
errors we have performed computer simulations.  For a 3D arrangement of six quantum
dots as shown in Fig.~\ref{fig:arch} (b), we randomly perturbed the positions of all
six quantum dots by up to four lattice sites in the $x$- and $y$-directions, and 
$\pm 1$ monolayer in the $z$-direction, assuming a lattice constant of $0.3$~nm,
values that appear feasible with current fabrication technology~\cite{PRL91n136104}, 
and numerically computed the average gate fidelity $\F$ for each perturbed system.  
Note that it was assumed here that the tunnelling rates can be kept steady despite
the effect of the silicon lattice by adjusting the control voltages appropriately.

The numerical results in Table~\ref{table:perturb} suggest that the robustness of 
our phase gate with respect to asymmetries depends significantly on the choice of
interqubit spacings and the distance between the auxiliary sites.  Concretely, the
data suggest that the robustness of the gate with respect to random pertubations of
the geometry is maximized by minimizing the distance between auxiliary sites while 
maximizing the distances between qubits.  There also appears to be a strong relation
between the robustness with respect to asymmetries and the \emph{effective} Coulomb
interaction between the auxiliary sites in the system, suggesting that maximizing 
the later quantity might be advantageous.  However, it should be noted that stronger
Coulomb coupling implies shorter gate operation times unless the gate parameters are
changed, which might affect the gate fidelity.  A final computation for a (target) 
geometry with $a=20$, $b=100$ and $c=10$ showed that the average gate fidelity over
100 random perturbations ranged from $0.9944$ to $1.0000$ with a mean (standard 
deviation) of $0.9989$ ($0.0013$).  These results suggest that even 3D charge-qubit 
architectures without shielding could be designed to be rather robust with respect 
to fabrication errors.

\begin{table}
%\rotatebox{90}{Fidelity}
\begin{tabular}{|*{5}{c|}} 
\hline
a $\backslash$ b & 30       & 50       & 70       & 90 \\\hline
20             & 0.964022 & 0.996631 & 0.998153 & 0.998855 \\\hline
40             & N/A      & 0.909071 & 0.992692 & 0.998410 \\\hline
60             & N/A      & N/A      & 0.831827 & 0.989618 \\\hline
80             & N/A      & N/A      & N/A      & 0.808333 \\\hline
\end{tabular} 
\caption{Mean of the average gate fidelity $\F$ for different geometries with fixed
intra-qubit spacing of $c=10$~nm.  For each target geometry the mean of $\F$ was 
computed for 30 randomly perturbed systems.  The gate parameters for all simulations 
were $n=k=1$ and the time steps were chosen such that the average gate fidelity of 
each \emph{unperturbed} systems was $\ge 0.99995$.  Notice that the mean of $\F$ 
increases sharply for decreasing values of $a$, and noticably for increasing values
of $b$.} \label{table:perturb}
\end{table}

%%%%%%%%%%%%%%%%%%%%%
\section{Conclusions}
%%%%%%%%%%%%%%%%%%%%%

We have presented two scalable achitectures for solid-state charge qubits that 
permit controlled entangling operations between pairs of qubits.  We add that 
these operations could also be performed in parallel for cluster state preparation,
a concept that we will explore elsewhere.  The key feature of both geometries is 
that interactions between qubits are mediated via auxiliary quantum dots, while 
direct interactions between qubits are suppressed either through the use of shielding, 
or a three-dimensional design to cancel interactions.  Controlled entangling gates
can therefore be implemented by switching the charge distribution, and hence the 
Coulomb interaction, between qubits using the auxiliary dots.  Both systems should
be realizable using existing or near-future fabrication techniques.

In particular, we have shown explicitly how to realize a controlled phase gate, 
i.e., a universal, maximally entangling two-qubit gate by a simple four-step 
procedure.  The crucial step in the sequence is the controlled tunnelling between
a qubit state and an auxiliary dot, which induces a phase shift conditional on the
occupation of an adjacent (auxiliary) quantum dot.
The scheme is sufficiently flexible to accommodate practical constraints on both
pulse lengths and tunnelling rates.  Analysis of the gate operation shows that the
controlled phase shift can be explained in terms of dynamic and geometric phases.
Due to the strength of the Coulomb coupling the gate operation is surprisingly fast.
In the absence of pulse length constraints gate operation times on the order of a
few picoseconds would be theoretically possible, and even with currently realistic 
constraints on the pulse lengths and tunnelling rates, gate operation times around
200 ps should be attainable.

Using computer simulations, we have also studied the effects of systematic errors
on the gate performance.  The simulations show that finite pulse rise and decay 
times tend to result in population loss to the auxiliary states, and can noticeably
reduce the average gate fidelity.  The average gate error increases sharply with 
increasing pulse rise and decay times.  The ideal scheme, however, can easily be 
generalized to compensate for such systematic errors.  Simulations show that these 
corrections can greatly improve the average gate fidelity; the corrected pulse 
sequences always achieved average gate fidelities of $>99.99$\%, and in some cases
the average gate fidelity increased from $\sim70$\% without corrections to $>99.99$\%.  
These results suggest that (experimental) characterization of the rise and decay 
times of the pulses is very important to achieve high gate fidelity.

We have also performed simulations to assess the robustness of the gate to noisy
control pulses.  The results suggest that the gate is quite robust with regard to
both bandwith-limited $1/f$ noise and white noise.  High-frequency noise tends to
effectively cancel.  Low-frequency fluctuations of the control pulses can reduce 
the gate fidelity.  Surprisingly, however, the resulting average gate errors are 
generally very small compared to the systematic errors.  In most of our simulations, 
the average gate error increased only from $<10^{-4}$ to $<10^{-3}$, even for very 
noisy controls.  Our simulations further suggest that even a 3D geometry without 
shielding, which relies mainly on symmetry to cancel the effect of the Coulomb 
interaction between qubits, can be made to be quite robust to misalignment errors
during fabrication if the parameters of the geometry are chosen sufficiently 
carefully.

\acknowledgments
The authors thank C.~J.~Wellard (Univ.\ of Melbourne) for useful discussions.  
S.G.S and D.K.L.O acknowledge financial support from the Cambridge-MIT Institute, 
Fujitsu and IST grants RESQ (IST-2001-37559) and TOPQIP (IST-2001-39215).  S.G.S 
also acknowledges support from the EPSRC IRC in quantum information processing, 
and D.K.L.O thanks Sidney Sussex College for support.  A.D.G is supported by the 
Australian Research Council, the Australian government and the US National Security
Agency (NSA), Advanced Research and Development Activity (ARDA) and the Army Research
Office (ARO) under contract number DAAD19-01-1-0653, and acknowledges support from
a Fujitsu visiting fellowship while visiting the University of Cambridge.

%%%%%%%%%
\appendix
%%%%%%%%%

%%%%%%%%%%%%%%%%%%%%%%%%%%
\section{Local Operations}
\label{app:local}
%%%%%%%%%%%%%%%%%%%%%%%%%%

We observed in Sec.~\ref{sec:local} that any local unitary operation can be realized 
by concatenating three elementary rotations.  It is worth noting, however, that some
local unitary operators can be implemented more efficiently if we allow simultaneous 
changes of more than one control parameter such as the energy difference $\Delta\eps_{12}
=(\eps_2-\eps_1)/2$ and the tunnelling rate $\mu_{12}$ between states $\ket{1}$ and 
$\ket{2}$.   To see this note that we can rewrite Eq.~(\ref{eq:Hloc2}) as follows:
\begin{equation}
  \op{H}(\eps_d,\mu_{12}) 
  = \bar{\eps}_{12}\op{I}_{12} + \Delta\eps_{12}\op{Z}_{12} 
    + \hbar\mu_{12}\op{X}_{12} + \eps_3\ket{3}\bra{3}
\end{equation}
where $\bar{\eps}_{12}=(\eps_1+\eps_2)/2$ and $\op{Z}_{12}=\ket{2}\bra{2}-\ket{1}\bra{1}$.
Thus, if we apply constant energy shifts $\eps_d$ and effect a fixed tunnelling rate 
$\mu_{12}$ between dots 1 and 2 (while all other tunnelling rates are kept zero) for 
$0\le t'\le t$ then we generate the unitary operator:
\begin{eqnarray}
 & & \exp[-it \op{H}(\eps_d,\mu_{12})/\hbar]  \nonumber\\
 &=& \left[\cos(\alpha t)\op{I}_{12} -i\frac{\sin(\alpha t)}{\alpha\hbar}
     (\Delta\eps_{12}\op{Z}_{12}+\hbar\mu_{12}\op{X}_{12})\right] \times \nonumber \\
 & & \exp(-it\bar{\eps}_{12}/\hbar) + \exp(-it\eps_3/\hbar) \ket{3}\bra{3} 
\end{eqnarray}
where $\alpha=\sqrt{(\Delta\eps_{12}/\hbar)^2+\mu_{12}^2}$.  Thus, we can realize a 
Hadamard transform, for instance, in a single step by setting $\Delta\eps_{12}=-
\hbar\mu_{12}\neq 0$ for time $t=\pi/(2\sqrt{2}\mu_{12})$.

%%%%%%%%%%%%%%%%%%%%%%%%%%%%%%
\section{First Swap Operation} 
\label{app:swap1}
%%%%%%%%%%%%%%%%%%%%%%%%%%%%%%

The total Hamiltonian for the first swap operation is $\op{H}_1=\op{H}_0+I_3\otimes
\op{H}^{(2)}$ where 
\begin{equation}
   \op{H}^{(2)} = \hbar\mu_{23}^{(2)} (\ket{2}\bra{3}+\ket{3}\bra{2})
\end{equation}
and $\op{H}_0=\ket{33}\bra{33}$ as before.  Applying this Hamiltonian for time $0\le t'
\le t$ gives rise to the unitary operator $\op{U}_1(t)=\exp[-(it/\hbar)\op{H}_1]$.   If
we set $t=\tau_1=\pi/(2\mu_{23}^{(2)})$ we obtain the block-diagonal matrix
\begin{equation}
  \op{U}(\tau_1) = \diag(1,-i X, 1, -i X, 1, W)
\end{equation}
where $X = \left( \begin{array}{cc} 0 & 1\\ 1 & 0 \end{array} \right)$ and $W$ 
is a symmetric $2 \times 2$ matrix with 
\begin{eqnarray*}
  W_{11} &=& e^{-i\tau_1/2} \left[\cos\left(\frac{\tau_1 u_1}{2}\right)
                        +\frac{i}{u_1}\sin\left(\frac{\tau_1 u_1}{2}\right)\right]\\
  W_{12} &=& e^{-i\tau_1/2} \left[-2i \frac{\mu_{23}^{(2)}}{u_1} 
                               \sin\left(\frac{\tau_1 u_1}{2}\right)\right]\\
  W_{22} &=& e^{-i\tau_1/2} \left[\cos\left(\frac{\tau_1 u_1}{2}\right)
                        -\frac{i}{u_1}\sin\left(\frac{\tau_1 u_1}{2}\right)\right]
\end{eqnarray*}
and $u_1=\sqrt{1+4[\mu_{23}^{(2)}]^2}$.  $\op{U}_1(t_1)$ swaps the population of 
states $\ket{2}_2$ and $\ket{3}_2$ provided that state $\ket{3}_1$ is not occupied.
%This assumption should normally hold but we could increase the robustness of the 
%gate by ensure that $W$ is diagonal 
%\begin{equation} 
%   W = (-1)^n \exp\left(-i \sqrt{4n^2-1} \pi/2\right) \diag(1,1).
%\end{equation}
%by setting 
%\begin{equation}
%  \mu_{23}^{(2)} =\frac{1}{2\sqrt{4n^2-1}}
%\end{equation}
%for some positive integer $n$.  

%%%%%%%%%%%%%%%%%%%%%%%%%%%%%%%
\section{Second Swap Operation}
\label{app:swap2}
%%%%%%%%%%%%%%%%%%%%%%%%%%%%%%%

The total Hamiltonian for the second gate is $\op{H}_2=\op{H}_0+\op{H}^{(1)}
\otimes I_3$ where 
\begin{equation}
   \op{H}^{(1)} = \frac{1}{2}\ket{2}\bra{2}
                  +\hbar\mu_{23}^{(1)}(\ket{2}\bra{3}+\ket{3}\bra{2})
\end{equation}
and $\op{H}_0=\ket{33}\bra{33}$ as before.  Applying this Hamiltonian for time 
$0\le t'\le t$ gives rise to the unitary operator $\op{U}_2(t)=\exp[-(it/\hbar)
\op{H}_2]$, which has the general form
\begin{equation}
   \op{U}_2(t) = \left(\begin{array}{c|cc} 
                  I_3 & 0 & 0 \\ \hline
                   0  & A & B \\
                   0  & B & C 
                 \end{array}\right)
\end{equation}
where $A=\diag(a,a,a')$, $B=\diag(b,b,b')$ and $C=\diag(c,c,c')$.  We must choose 
the gate operation time $\tau_2$ such that $B=0$, i.e., $b=b'=0$.  Since
\begin{eqnarray}
  b  &=& -4i \exp\left(\frac{-i t}{4} \right) 
          \frac{\mu_{23}^{(1)}}{u_2} \sin\left(\frac{u_2 t}{4} \right) \\
  b' &=& -4i \exp\left(\frac{-3it}{4} \right) 
          \frac{\mu_{23}^{(1)}}{u_2} \sin\left(\frac{u_2 t}{4} \right) 
\end{eqnarray}
with $u_2=\sqrt{1+16[\mu_{23}^{(1)}]^2}$, this is equivalent to $u_2 T_2= 4n\pi$ 
for some integer $n$, or
\begin{equation}
   \tau_2 = \frac{4\pi n}{\sqrt{1+ 16[\mu_{23}^{(1)}]^2}}.
\end{equation}
This choice of gate operation time gives 
\begin{equation}
   U_2(\tau_2) = \diag(1,1,1,a,a,a',a,a,a')
\end{equation}
where we have 
\begin{equation} \label{eq:a}
  \begin{array}{rcl}
  a  &=& (-1)^n \exp(-i n\pi/u_2), \\ 
  a' &=& (-1)^n \exp(-i 3n\pi/u_2).  
  \end{array}
\end{equation}

%%%%%%%%%%%%%%%%%%%%%%%%%%%%%%%
\section{Local Phase Rotations}
\label{app:step4}
%%%%%%%%%%%%%%%%%%%%%%%%%%%%%%%

Composing the three swap operations, noting $\op{U}_3(\tau_3)=\op{U}_1(\tau_1)$, gives
\begin{eqnarray}
  \op{U} &=& \op{U}_1(\tau_1) \op{U}_2(\tau_2) \op{U}_1(\tau_1)  \nonumber \\  
         &=& \diag(1,-1,-1,a,-a',-a,a,W \diag(a,a') W)  \nonumber \\ 
\end{eqnarray}
whose projection onto the two-qubit subspace $\{\ket{11},\ket{12},\ket{21},
\ket{22}\}$ is $\tilde{U}=\diag(1,-1,a,-a')$ with $a$ and $a$ as in (\ref{eq:a}),
which is not quite a controlled phase gate yet.  However, if we multiply $\tilde{U}$ 
by $U^{(1)} \otimes U^{(2)}$, where
\begin{equation}
  \begin{array}{rcl}
  U^{(1)} &=& \diag\left(1,(-1)^n \exp(i\pi n/u_2) \right), \\
  U^{(2)} &=& \diag(1,-1)
  \end{array}
\end{equation}
which corresponds to simultaneous local phase rotations on state $\ket{2}_1$ and 
$\ket{2}_2$, we obtain
\begin{equation}
  U^{(1)} \otimes U^{(2)} \tilde{U} = \diag\left(1,1,1,\exp(-i 2\pi n/u_2) \right)
\end{equation}
where $u_2=\sqrt{1+16[\mu_{23}^{(1)}]^2}$ and
\begin{equation}
  \mu_{23}^{(1)} = \frac{1}{4} \sqrt{\left(\frac{2n}{2k-1}\right)^2-1}
\end{equation}
for positive integers $n$ and $k$ satisfying $2n>2k-1$.  Then clearly $2\pi n = 
(2k-1)\pi u_2$, i.e., we have
\begin{equation}
   U^{(1)} \otimes U^{(2)} \tilde{U} = \diag(1,1,1,-1) = \op{U}_{\rm phase}.
\end{equation}
To achieve a controlled phase gate on the qubit space we thus require
\begin{equation}
  \exp(-i2\pi n/u_2) = -1 \quad \Leftrightarrow \quad 2\pi n = (2k-1)\pi u_2.
\end{equation}
Substituting $u_2=2n/(2k-1)$ yields $(-1)^n e^{i\pi n/u_2}=e^{i\pi(n+k-1/2)}$, 
which shows that $\op{U}^{(1)}=\op{U}_2^{(1)}(\ell\pi)$ with $\ell=1/2-(n+k)
\mbox{ mod }2$ and $\op{U}^{(2)}=\op{U}_2^{(2)}(\pi)$, i.e., we must set 
$\eps_2^{(1)}=\ell\pi/\tau_4$ and $\eps_2^{(2)}=\pi/\tau_4$ for some time 
$t=\tau_4$ in the final step.

%%%%%%%%%%%%%%%%%%%%%%%%%%%%%%%%%%%%%%%%%%%%%%%%%%%%%%
\section{Tunnelling Rate and Pulse Length Constraints} 
\label{app:constraints}
%%%%%%%%%%%%%%%%%%%%%%%%%%%%%%%%%%%%%%%%%%%%%%%%%%%%%%

In absence of constraints on the tunnelling rate $\mu_{23}^{(1)}$ it is easy to 
see that the optimal choice of the parameters $n$ and $k$ in the second step is
$k=n=1$, which implies $u_2=2$ and $\mu_{23}^{(1)}=\sqrt{3}/4$.  In practice, 
however, the control pulses usually cannot be arbitrarily short and there the 
tunnelling rates cannot be made arbitrarily large, and we can accommodate these 
constraints by choosing suitable values for the parameters $n$ and $k$.

If the mimimum pulse length is $\tau_2\ge T_{min}$ but the tunnelling rates are 
unconstrained then we simply set
\begin{equation}
   k = \ceil{\frac{\tau_{min}}{4\pi}+\frac{1}{2}}, \quad n=k.\label{eq:kn}
\end{equation}     
where $\ceil{x}$ is the smallest positive integer $\ge x$, to optimize the gate
operation time while satisfying the pulse length constraint.

If we must satisfy $\mu_{23}^{(1)}\le\mu_{max}$ but there is no constraint on
the length of the control pulses, we set
\begin{eqnarray}
   k &=& \floor{\ceil{(u_{max}-1)^{-1}}/2} +1 \label{eq:k1}\\
   n &=& \floor{(2k-1) u_{max}/2}             \label{eq:n1}
\end{eqnarray}
where $u_{max}=\sqrt{16\mu_{max}^2+1}$ and $\floor{x}$ is the largest (positive)
integer $\le x$, to satisfy the constraint and optimize the gate operation time.

Finally, if we must satisfy $\tau_2 \ge \tau_{min}$ and $\mu_{23}^{(1)}\le\mu_{max}$,
we choose
\begin{eqnarray}
   k &=& \max\left\{k_1,\ceil{\tau_{min}/4\pi+1/2} \right\} \label{eq:k2}\\
   n &=& \floor{(2k-1) u_{max}/2}                        \label{eq:n2}
\end{eqnarray}
where $k_1=\floor{\ceil{(u_{max}-1)^{-1}}/2}+1$ as in (\ref{eq:k1}) and $u_{max}=
\sqrt{16\mu_{max}^2+1}$ as before.

\bibliography{papers,sonia,papers_old,books,PhaseGateRefs}
\bibliographystyle{prsty}
\end{document}